\documentclass[fleqn,usenatbib]{mnras}

\usepackage{newtxtext,newtxmath}

\usepackage[T1]{fontenc}

\DeclareRobustCommand{\VAN}[3]{#2}
\let\VANthebibliography\thebibliography
\def\thebibliography{\DeclareRobustCommand{\VAN}[3]{##3}\VANthebibliography}

\usepackage{graphicx}	
\usepackage{amsmath}	
\usepackage{pdflscape}  

\newcommand{\snum}[1]{#1}
\newcommand{\tnum}[1]{#1}

\title[X-ray--radio multi-dimensional LF]{Exploring the X-ray--radio connection for AGN via measurements of the multi-dimensional luminosity function}

\author[C. M. Pennock et al.]{
Clara M. Pennock,$^{1}$\thanks{E-mail: clara.pennock@ed.ac.uk }
James Aird,$^{1}$
Cassandra L. Barlow-Hall,$^{1}$
\\
Institute for Astronomy, University of Edinburgh, Royal Observatory, Blackford Hill, Edinburgh EH9 3HJ, UK$^{1}$\\
}

\date{Accepted XXX. Received YYY; in original form ZZZ}

\pubyear{2025}

\begin{document}
\label{firstpage}
\pagerange{\pageref{firstpage}--\pageref{lastpage}}
\maketitle

\begin{abstract}
We present new methods to quantify the AGN population in terms of a multi-dimensional luminosity function that describes the space density of sources as a function of both X-ray and radio luminosity. 
We compile a sample of \tnum{1538} radio and X-ray detected extragalactic sources from the Bo\"otes and COSMOS fields. First, we investigate the X-ray--radio luminosity correlation in the sample
and find that an apparent correlation is introduced due to the sensitivity limits of the surveys; when considering individual redshift bins we find a wide range of radio luminosities associated with a given X-ray luminosity, and vice versa, indicating little direct connection between the emission processes.
We then measure the X-ray luminosity function, radio luminosity function and multi-dimensional X-ray--radio luminosity function 
across redshift ($0<z<6$). 
We apply luminosity thresholds in X-ray and radio to restrict our sample to those in the AGN-dominated regime and explore how the fraction of radio-selected AGN within the overall X-ray sample varies with increasing X-ray luminosity (and vice versa). We find that towards the highest X-ray and radio luminosities the fraction of sources with both an X-ray and radio detection increases towards 100\%, indicating that at the highest luminosities we are more likely to obtain a detection in both bands, though the source will not necessarily be bright in both bands.
Thus, the most luminous accretion events are more likely to be associated with the production of a jet, despite the distinct physical structures that produce the emission and likely persist over very different timescales.


\end{abstract}




\section{Introduction}

Actively accreting supermassive black holes (SMBH), also known as Active Galactic Nuclei (AGN), can produce radiation across the electromagnetic spectrum \citep[e.g.][and references therein]{Padovani2017}. The source of the emission in the AGN structure is distinct to each wavelength, such as the accretion disk which radiates in the UV/optical and the dusty obscuring structure, often referred to as the dusty torus, which radiates in the infrared. The primary originator of X-ray emission in an AGN is the inverse Compton scattering of photons from the accretion disc, which takes place in a hot corona. For the brightest radio emitters, often referred to as the radio-loud population which are expected to only take up $\sim$10\% of the entire AGN population, the emission is primarily from synchrotron emission in large (up to megaparsec scales) powerful relativistic jets. The origin of radio emission from the more numerous, radio-quiet sample is less clear and could be from a corona, a small-scale version of relativistic jets, AGN-driven winds or even star-formation, in which case the emission would not be directly associated with the AGN \citep[see ][and references therein]{Panessa2019}.

Radio and X-ray emission are often good indicators of the presence of an AGN, even when the AGN is obscured at other wavelengths \citep[see][and references therein]{Hickox2018}. The detection of X-ray emission is a reliable indicator due to X-ray emission from other astrophysical processes (e.g. X-ray binaries) being typically weak in comparison, and only the most massive and highly star-forming (and rarer) galaxies being capable of producing X-ray emission at $L_{\mathrm{X}}>10^{42}$ erg s$^{-1}$ (where AGN are expected to dominate). Therefore host galaxy contamination is only an issue for low luminosity AGN and heavily obscured AGN, where the X-ray emission is suppressed via absorption. 

Similarly, in the radio, whilst star-forming galaxies can be bright in radio, AGN are easily distinguishable from star-formation at the highest radio luminosities (e.g. $>10^{25}$~W~Hz$^{-1}$). Furthermore, there is a known tight correlation between the expected radio and far-IR luminosities for star-formation processes \citep[e.g.][]{Bell2003,Ivison2010,Heeson2022}, which can be used to identify sources that have excess radio emission over that expected from star-formation, and are therefore AGN \citep[e.g.][]{Donley2005, Delvecchio2017, Best2023}. However, it should be noted that the radio-excess method will fail to identify low-to-moderate radio luminosity AGN in galaxies with substantial star-formation, rendering the selection incomplete. Radio also has the benefit of having very low optical depth, and therefore radio selection can be used to identify the most obscured sources \citep[e.g.][]{Hickox2018}. There are several dichotomies that have been observed in radio populations amongst which is radiative-mode vs jet-mode AGN \citep{Best2012, Heckman2014}. In a radiative-mode AGN there is a geometrically-thin, optically-thick accretion disk that undergoes radiatively efficient accretion. In radiative-mode we observe the signature of an AGN in multiple bands, where what we observe is dependent on our line of sight (e.g. Type 1 vs Type 2) and there is the possibility of a powerful radio jet being produced. In contrast, in a jet-mode AGN the accretion is radiatively inefficient, there is an advection-dominated accretion flow which replaces the inner part of the accretion disk (which is either truncated or entirely absent), and we only observe an AGN signature in the radio, predominantly as radio jets. The X-ray emission in a radiative-mode AGN are produced by the hot corona from inverse Compton scattering of photons from the accretion disk, however, in jet-mode AGN the accretion disk is either truncated or entirely absent, leading to weaker X-ray emission.

Multi-wavelength surveys in fields such as Bo\"otes \citep[][]{2021Tasse,2020Masini}, COSMOS \citep{2017bSmolcic,2016Civano}, ELAIS-S1 \citep{Franzen2015, Ni2021} and XMM-LSS \citep{Heywood2020, Chen2018}, that have been observed by both radio (e.g. VLA\footnote{Very Large Array}, ATCA\footnote{Australia Telescope Compact Array}, LOFAR\footnote{LOw Frequency ARray}, ASKAP\footnote{Australian Square Kilometre Array Pathfinder}) and X-ray (e.g \textit{Chandra}, \textit{XMM-Newton}, \textit{Swift}) telescopes have been a large boon for the many investigations into the relationship between the X-ray and radio emission of AGN. The `Fundamental Plane of black hole activity' \citep[e.g. ][]{2003Merloni, Falcke2004} is a well-known correlation that provides a link between the SMBH mass, accretion disk (traced by the X-ray emission) and a relativistic jet (traced by the radio emission). X-ray and radio emission have been found to follow a linear correlation (in logarithmic space) with steeper slopes being attributed to bright AGN which display a radiatively efficient accretion mechanism \citep[e.g.][]{Coriat2011, Dong2014}, whilst others show a shallower slope which has been linked to low luminosity AGN that exhibit radiatively inefficient accretion \citep[e.g. ][]{2003Gallo, 2003Merloni, dAmato2022}. However, there is generally a large scatter around these relations ($>$1--2~dex), and it has been proposed that the correlations are the result of selection bias \citep[e.g.][]{Mingo2014}.

Rather than quantifying the relationships within observed samples, measurements of the luminosity functions of AGN in both the X-ray and radio bands provide a means of quantifying the intrinsic space density of sources across differing luminosities and over cosmic time, capturing the diversity of the AGN population. 
The AGN X-ray luminosity function (XLF) can be best fit by a double power-law which evolves with redshift according to a luminosity-dependent density evolution (LDDE) model \citep[e.g. ][]{Ueda2014, Aird2015, Pouliasis2024} and the AGN radio luminosity function (RLF) can also be fit with a double power law that evolves with redshift \citep[e.g.][]{Mauch2007,Smolcic2017c,Novak2018} and may be described by either pure luminosity evolution (PLE) or pure density evolution (PDE) models. Luminosity functions (LFs), however, tend to only look at the space densities in a single waveband, and inclusion of other wavelengths would be as subsets of varying properties in the additional waveband. An LF in more than one waveband, a multi-dimensional LF, has not been explored yet\footnote{Except in \citet{1983A&A...125..223V}, where only a sample of 104 cD galaxies was investigated in the radio, X-ray and optical.}. Exploring a multi-dimensional LF would allow us investigate how AGN evolve in the X-ray and radio concurrently, as well as to see if there are any links/correlations between the emission produced at different distances to the SMBH, corona/accretion disk (X-ray) and relativistic jets (radio), across cosmic time.

In this paper, after defining our sample taken from the Bo\"otes and COSMOS fields in Section \ref{Sec:Data}, we first investigate the X-ray--radio correlation, whilst taking into account survey sensitivity limits, by looking at the X-ray and radio detected sources in the Bo\"otes and COSMOS fields in Section \ref{sec:LXLRcomparison}. Then, in Section \ref{method} we define a method for calculating a luminosity function that estimates the space density of sources in both X-ray and radio luminosity space. Next, in Section \ref{sec:LF}, we apply this to the data and explore the resulting luminosity functions. Furthermore, in Section \ref{sec:AGNdom} we restrict the X-ray and radio detected sample to an AGN dominant population and investigate how the fraction of X-ray and radio detected AGN to the overall X-ray (and radio) population changes with increasing X-ray and radio luminosity. In Section \ref{sec:disc} and \ref{sec:conc} we discuss the results and summarise the conclusions, respectively. 

Throughout this work we adopt a Flat $\Lambda$CDM cosmology with $H_0$ = 70~km s$^{-1}$ Mpc$^{-1}$, $\Omega_M$ = 0.3 and $\Omega_\Lambda$ = 0.7.

\section{Data and sample definition}\label{Sec:Data}

For this study we require fields that have been observed both in the X-ray and the radio. We also require them to be observed in the optical/IR, from which photometric redshifts can be measured from modelling SEDs. We utilise the Bo\"otes field (Section \ref{BootesData}) which is a wide field, and the COSMOS field (Section \ref{COSMOSData}), which is a smaller but deeper field. Both these fields also have benefited from complementary dedicated spectroscopic campaigns from which redshifts can be drawn.
\subsection{COSMOS}\label{COSMOSData}

The Cosmic Evolution Survey \citep[COSMOS;][]{2007COSMOS}\footnote{http://cosmos.astro.caltech.edu/} is a survey over a 2 deg$^{2}$ area that includes multi-wavelength imaging from X-ray to radio as well as extensive optical spectroscopy.

The UltraVISTA \citep{UltraVISTA2012} survey covered a 1.5 deg$^{2}$ area within the COSMOS field in the near-IR, and provided a deeper survey from which to detect higher-z objects and improve photometric redshift estimates. In this study we concentrate on sources in the area provided by the UltraVISTA survey. 

\subsubsection{Spectroscopic redshifts}\label{COS-spec}
There have been several spectroscopy campaigns for the COSMOS field\footnote{A summary of which can be found in Table \ref{COSspectab}.}$^,$\footnote{In the course of this work a catalogue of spectroscopy in the COSMOS field has been compiled and made public \citep{COSMOSspec}. This catalogue was used to gain new/improved spectroscopic redshifts for 607 radio and 162 X-ray detected sources.}. These include zCOSMOS \citep{2007Lilly}, with the most recent data release, zCOSMOS-bright DR3 \citep{Lilly2023}, including spectra for 20,689 objects across 1.7 deg$^{2}$ of the COSMOS field. After we applied the recommended redshift reliability flags (which are generally based on the number of features, such as emission/absorption lines, that can be reliably identified) from \citet{2007Lilly}, 17,980 sources remained. The Magellan telescope was used to observe X-ray and radio-selected AGN targets \citep[][]{2007Magellan}, observing a total of 1338 sources, 771 of which have good spectra quality. PRIsm MUlti-object Survey \citep[PRIMUS;][]{2011PRIMUS,2013PRIMUS}, a low-resolution spectroscopic survey, obtained redshifts for $\sim$10,000 sources in the COSMOS field. A Wide-field Grism Spectroscopic Survey with the Hubble Space Telescope \citep[3D-HST;][]{3DHST2012,3DHST2016} observed 33,879 sources in the COSMOS field, 4158 of which have a redshift. VIMOS Ultra Deep survey \citep[VUDS;][]{2015VUDS}, which contains 291/394 sources with a reliable redshift after we applied the recommended cuts using the given redshift reliability flags in \citet{2015VUDS}. DEIMOS 10K Spectroscopic Survey Catalog of the COSMOS Field \citep[][]{2015DEIMOS}, which contains 7988/10,718 sources with a reliable redshift, after we applied recommended cuts using redshift reliability flags from \citet{2015DEIMOS}. The MOSFIRE Deep Evolution Field Survey \citep[MOSDEF;][]{2015MOSDEF} observed 616 sources in the COSMOS field, 431 sources of which have good quality redshifts. The Fiber Multi-Object Spectrograph (FMOS)-COSMOS survey \citep{2019FMOS} contains 5484 sources, 1931 of which have a redshift. SDSS \citep[SDSS DR18;][]{2023SDSSDR18} spectroscopically observed $\sim$1530 sources in the COSMOS field, and there is a specific SDSS Quasar Catalogue \citep[SDSS DR16Q;][]{2020SDSSDR16Q}, which has $\sim$170 sources in the COSMOS field. Further sources were observed with optical spectroscopic campaigns specifically targeting X-ray sources identified in the XMM-COSMOS \citep[931 sources;][]{2010XMMz},
the C-COSMOS survey \citep[901 sources;][]{2012Civano} and the
C-COSMOS Legacy survey \citep[1761 sources;][]{2016Marchesi}.

We combined these catalogues with a matching radius of 1$^{\prime\prime}$, and where there is overlap/duplicates and the redshifts disagree, we selected the more confident redshift, based on redshift reliability flags from the catalogue each source is from, as the `best' redshift (where redshift reliability flags were the same, the observation with the better spectral resolution was used). The combined catalogue contains 36,564 spectroscopically observed sources in the COSMOS field. A summary of the spectroscopic catalogues can be found in Table \ref{COSspectab}, where we also note how many of these sources were radio or X-ray detected.

\begin{table}
\caption{Summary of spectroscopic catalogues used in the COSMOS field, in order of preference (based on, e.g. spectral resolution of the survey). The `X-ray sources' and `Radio sources' columns give the number of spectroscopic redshifts that we adopt in our study for each sample from the different catalogues. If there is a spectroscopic redshift from more than one of these catalogues, then the redshift we adopt will be taken from the catalogue that is higher up the list.}
    \centering
    \begin{tabular}{lcccl}
        \hline\hline
        \llap{C}atalogue & \llap{To}tal no. & X-ray & Radio & Ref.\\
        & \llap{so}urces & \llap{s}ource\rlap{s} & \llap{s}ource\rlap{s} &\\
        \hline
        \llap{M}OSDEF & 431 & ~~10 & ~~13 & \citet{2015MOSDEF}\\
        \llap{V}UDS & 263 & ~~~~1 & ~~~~0 & \citet{2015VUDS}\\
        \llap{F}MOS & \llap{1}909 & 198 & 407 & \citet{2019FMOS}\\
        \llap{D}EIMOS & \llap{7}772 & 463 & 848 & \citet{2015DEIMOS}\\
        \llap{C}-COSMOS Lega\rlap{cy} & \llap{1}223 & 753 & 514 & \citet{2016Marchesi}\\
        \llap{C}-COSMOS & ~~13 & ~~~~8 & ~~~~1 & \citet{2012Civano}\\
        \llap{X}MM-COSMOS & 100 & ~~~~6 & ~~16 & \citet{2010XMMz}\\
        \llap{M}agellan & 288 & ~~~~5 & 213 & \citet{2007Magellan}\\
        \llap{z}COSMOS-bright & \llap{15,}793 & ~~11 & \llap{1}007 & \citet{2007Lilly}\\
        \llap{S}DSS DR16Q & ~~57 & ~~~~0 & ~~~~1 & \citet{2020SDSSDR16Q}\\
        \llap{S}DSS DR18 & 982 & ~~~~1 & ~~31 & \citet{2023SDSSDR18}\\
        \llap{3}DHST & \llap{3}005 & ~~~~7 & ~~59 & \citet{3DHST2012}\\
        \llap{P}RIMUS & \llap{4}728 & ~~~~3 & 210 & \citet{2011PRIMUS}\\
        \hline
    \end{tabular}
    \label{COSspectab}
\end{table}

\subsubsection{Radio}
The COSMOS field was observed in the radio as part of the \textit{Karl G. Jansky} Very Large Array Cosmic Evolution Survey (VLA-COSMOS) 3~GHz Large Project \citep{2017aSmolcic}. This is a radio continuum survey performed at 10 cm and covering a 2.6 deg$^{2}$ area, encompassing the entire COSMOS field, with a mean rms of 2.3 $\sim$ $\mu$Jy beam$^{-1}$. From this a catalogue of \snum{10,830} radio sources with 5$\sigma$ detections was created.

The optical/near-IR counterparts to the radio source catalogue were found by \citet{2017bSmolcic} by primarily cross-matching with the photometry and photometric redshift catalogue from \citet{COSMOS2015}, 
which contains optical and NIR photometry in over 30 bands for \snum{1,182,108} sources over a $\sim$2.3 deg$^{2}$ area. When regions containing bright stars are masked out, the area reduces to 1.77 deg$^{2}$, providing a catalogue of 9191 radio sources with optical/NIR counterparts.

A more recent COSMOS photometry and photometric redshift catalogue \citep{COSMOS2020} has since been released. We take the counterparts identified at optical/near-IR/mid-IR wavelengths in \citet{2017bSmolcic} and cross-match to the \citet{COSMOS2020} catalogue, providing updated photometry for all the radio sources with counterparts in the \citet{2017bSmolcic} catalogue. We also searched for additional counterparts for sources that lack a cross-identification in \citet{2017bSmolcic} by cross-matching the radio source positions with the FARMER COSMOS2020 catalogue of \citet{COSMOS2020} with a matching radius of 0.6$^{\prime\prime}$ \citep[as used in][]{2017bSmolcic}, providing counterparts for an additional \snum{1808} radio sources. For sources with no spectroscopy, we use the photometric redshifts from \citet{2016Marchesi} where the radio source has an associated X-ray source, and where these are not available we restrict to sources  where the reduced $\chi^2 <$ 10 for the SED model fit to retain reliable photometric redshifts, as well as choose the redshift with the model fit (AGN, galaxy, star) with the lowest reduced $\chi^2$.

We used the COSMOS FARMER catalogue, restricted to the \tnum{7218} radio sources in the UltraVISTA survey area, \tnum{6599} of which have a counterpart. 
For this paper, we also imposed a 3~GHz flux limit of 11.5 $\mu$Jy, which is the 5$\sigma$ level at an angular resolution of 0.75$^{\prime\prime}$ \citep{2017aSmolcic}, below which source counts start to drop and the catalogue is not complete.
Cross-matching with the COSMOS photometry/spectroscopy yielded \tnum{6210} sources above our radio flux limit that have a redshift measurement of $z>$0 (spectroscopic or photometric), \tnum{3107/6210} ($\sim 50$\%) of which are spectroscopic redshifts. 

\subsubsection{X-ray}
The COSMOS field was observed in the X-ray by \textit{Chandra} as part of the COSMOS-Legacy survey \citep{2016Civano}, which imaged 2.2 deg$^{2}$, reaching an effective exposure of $\sim$ 160 ks and $\sim$ 80 ks over the central 1.5 deg$^{2}$ and remaining area, respectively. This survey combined 56 new observations with the previous C-COSMOS survey.
In this study we utilise an X-ray source catalogue obtained from the \textit{Chandra} COSMOS-Legacy Survey data \citep{2016Civano} through a custom data reduction and source detection process first detailed in \citet{Laird2009}, the resulting source catalogue of which has been used in previous works \citep{Aird2017, Aird2018, Aird2019, Laloux2023, Laloux2024, BarlowHall2025}.

This X-ray sample was limited to the \tnum{2408} sources lying in the UltraVista region. A cross-match to the FARMER COSMOS2020 catalogue of \citet{COSMOS2020} was performed using the Bayesian cross-matching code NWay \citep[][]{Buchner2021}, which considers both positional offsets and the expected multiwavelength properties when assessing the probability that a given object is the true counterpart to an X-ray source.
Here, we adopt internally generated priors for the $K_\mathrm{S}$-band magnitudes and IRAC ch1-ch2 colours, which are based on an initial (conservative) cross-matching. This process is detailed in \citet{BarlowHall2025} and provides counterparts for \tnum{2388/2408 (99\%)} of the COSMOS sources. 
For sources with no spectroscopy, we use the photometric redshifts from \citet{2016Marchesi}, and where these are not available we adopt the photo-z from \citet{COSMOS2020} but restrict to sources where the reduced $\chi^2 <$ 10 for the SED model fit to retain reliable photometric redshifts, as well as choosing the redshift from the model fit (AGN, galaxy, star) with the lowest reduced $\chi^2$.
For this paper, we select sources that are detected in the full (0.5--7~keV energy band) to a false probability threshold $<4\times10^{-6}$ and have redshifts $z>0$.
This provides a sample of \tnum{2158} X-ray sources, \tnum{1481} with a spectroscopic redshift measurement.

In X-ray surveys the sensitivity across the survey area is not uniform. A sensitivity curve is the completeness of a survey as a function of flux and provides a detection probability for a source of a given flux. We derive the sensitivity curve and then rescale by the total area of the survey, creating a sensitivity curve that can be used to correct for survey completeness.
For our COSMOS sample, we use the sensitivity curves that were derived by \citet{BarlowHall2025}, following the method of \citet{Georgakakis2008}.

\subsection{Bo\"otes}\label{BootesData}
The Bo\"otes field is a 9.3 deg$^{2}$ area of sky that was originally observed as part of the optical NOAO Deep Wide Field Survey \citep[NDWFS][]{1999NDWFS}. Since then it has been extensively observed multiple times from X-ray to radio.

\subsubsection{Spectroscopic and photometric redshifts}

The sources with known spectroscopy in the Bo\"otes field have been collated as part of the multi-wavelength photometric catalogue created by \citet{2021Duncan} that contains both photometric and spectroscopic redshifts. The spectroscopic redshifts account for \snum{21,857} out of the total \snum{2,214,329} sources in the photometric catalogue, and are primarily from the AGN and Galaxy Evolution Survey \citep[AGES;][]{AGES}.
\subsubsection{Radio}
 
In the radio band, the Bo\"otes field was observed as part of the LOFAR Two Metre Sky Survey (LoTSS): the LoTSS Deep Fields \citep{2021Tasse}. It was observed at frequencies between 114.9--177.4 MHz down to sensitivities of around 32 $\mu$Jy beam$^{-1}$ over an area of $\sim$9.2 deg$^{2}$. Source extraction on the radio images was performed to create a catalogue of \snum{36,767} radio detections which were then cross-matched with their optical counterparts in \citet{2021Kondapally}, by matching the radio detections to the multi-wavelength catalogue created by \citet{2021Duncan} that contains both photometric and spectroscopic redshifts. The overlap between the radio and the multi-wavelength coverage \citep[defined in Sect. 2 of][]{2021Kondapally} reduced the area to $\sim$8.63 deg$^{2}$ and the number of radio sources to \snum{19,179} of which \tnum{18,579} have counterparts.

We remove sources from areas that are masked due to bright stars in optical/IR surveys by applying flags\footnote{we require that `FLAG\textunderscore CLEAN' is set to 0 in the \citet{2021Kondapally} catalogue.}. We also limit to sources with radio detections with peak flux density at $>$5$\sigma$ level based on local rms. We also imposed an additional limit of flux density, $F_{\rm 140~MHz}$ $>$ 32 $\mu$Jy, as this is where the source counts start to drop in the distribution of flux densities, and is the RMS noise level of the Bo\"otes image, below which the radio catalogue becomes incomplete. 
This leaves \tnum{12,670} sources above our radio flux limit which have $z>$0, \tnum{3255/12,670} of which have spectroscopic redshifts. 

\subsubsection{X-ray}

In the X-ray band, the Bo\"otes field was observed with the \textit{Chandra} X-ray Observatory as part of the 
\textit{Chandra} Deep Wide-Field Survey \citep[CDWFS;][]{2020Masini}. CDWFS combined data from 281 \textit{Chandra} pointings in the Bo\"otes field conducted between 2003 and 2018, for a total exposure time of 3.4 Ms over an area of $\sim$9.3 deg$^{2}$. From which a point source catalogue was created of \tnum{6891} sources down to limiting fluxes of 4.7 $\times$ 10$^{-16}$, 1.5 $\times$ 10$^{-16}$ and 9 $\times$ 10$^{-16}$ erg cm$^{-2}$s$^{-1}$ for the full (0.5--7~keV), soft (0.5--2~keV) and hard (2--7~keV) bands, respectively. Spurious sources are removed based on a probability threshold $P$, where the probability of a source being real (i.e. not spurious) increases below this threshold, where $\log P$ = -4.63, -4.57, -4.40, in the full, soft and hard bands. In this paper we only consider the sources detected in the full band.

The X-ray point source catalogue was matched to the multi-wavelength catalogue created in \citet{2021Duncan} using the catalogue IDs from \citet{2021Duncan} \citep[see][for more information on how these were cross-matched]{2020Masini}, and it is from here that the redshifts are taken. From this catalogue we select \tnum{5333/6891} of the X-ray sources that are detected in the full band, have counterparts, and have a redshift $z>0$. \tnum{2397/5333} of these redshifts are spectroscopic and the remaining are photometric.

To account for completeness in the Bo\"otes field we use \textit{Chandra} sensitivity curves derived by  \citet{2020Masini} based on the methods of \citet{Georgakakis2008} that were confirmed via simulations \citep[as described in][]{2020Masini}.

\subsection{X-ray--radio sample}

For the COSMOS sample we cross-matched the X-ray and radio catalogues via their COSMOS2015/COSMOS2020 catalogue IDs. This gave \tnum{639} matches, of which \tnum{600} have $z >$ 0, \tnum{449} of which have a spectroscopic redshift. 

For the Bo\"otes sample, due to the shared matching of the radio and X-ray catalogues with the catalogue created in \citet{2021Duncan}, matching the radio to the X-ray was straightforward, and simply required matching to the source IDs from the multi-wavelength catalogue. This gave \tnum{1255} matches, \tnum{938} of which have $z$ $>$ 0, \tnum{492/938} of which are spectroscopic.

A summary of the number of sources for the Bo\"otes and COSMOS fields for radio, X-ray and X-ray+radio that have spectroscopic and photometric redshifts can be found in Table \ref{Tab:redshift}. To assess the reliability of the photometric redshifts we used the standard metrics for measuring photometric redshift quality/reliability \citep[e.g. see][]{Salvato2022}. These include the outlier fraction, $\eta$, defined as the fraction of sources for which $|z_{\rm phot}-z_{\rm spec}|/(1+z_{\rm spec})>0.15$ \citep[e.g. ][]{Hildebrandt2010, Salvato2022}, as well as the accuracy, $\sigma_{\rm NMAD}$, defined as 1.48$\times$median($|z_{\rm phot}-z_{\rm spec}|/(1+z_{\rm spec})$) \citep{Ilbert2006}. The values for $\eta$ and $\sigma_{\rm NMAD}$ for each survey and sample (X-ray, radio, X-ray+Radio) can be found in Table \ref{Tab:redshift}.

\begin{table*}
\caption{Summary of the counts of the sources detected within the considered areas of the COSMOS and Bo\"otes fields, separated into waveband and the number of sources that have optical/IR counterparts (CTP column). The ``Spec-z'' and ``Phot-z'' columns give the number of sources that enter our final samples (after applying flux limits or detection criteria, see Section~\ref{Sec:Data}) that have either a spectroscopic redshift or a photometric redshift (adopted when no spectroscopic redshift is available) of $z>0$ \citep[restricted to reliable redshifts and including updated redshifts from][ in the COSMOS field]{COSMOSspec}. Also, to assess the photometric redshifts, the outlier fraction, $\eta$, defined as the fraction of sources for which $|z_{\rm phot}-z_{\rm spec}|/(1+z_{\rm spec})>0.15$ \citep[e.g. ][]{Hildebrandt2010, Salvato2022} and the accuracy, $\sigma_{\rm NMAD}$, defined as 1.48$\times$median($|z_{\rm phot}-z_{\rm spec}|/(1+z_{\rm spec})$) \citep{Ilbert2006}, is also given.}
    \centering
    \begin{tabular}{lcccccccccccc}
        \hline\hline
        Detection & \multicolumn{6}{c}{------------------------ COSMOS ------------------------} & \multicolumn{6}{c}{---------------------------- Bo\"otes ----------------------------} \\
        Band & All & CTP & Spec-z & Phot-z & $\eta$ & $\sigma_{\rm NMAD}$ & All & CTP & Spec-z & Phot-z & $\eta$ & $\sigma_{\rm NMAD}$ \\
        \hline
        X-ray & 2408 & 2388 & 1481 & ~~677 & 9.51\% & 0.024 & ~~~6891 & ~~~6864 & 2397 & ~~~2936 & 14.04\% & 0.056\\
        Radio & 7218 & 6599 & 3107 & 3103 & 6.34\% & 0.020 & 19,179 & 18,579 & 3255 & ~~~9415 & ~~4.09\% & 0.019\\
        X-ray+Radio & - & ~~639 & ~~449 & ~~151 & 8.87\% & 0.024 & - & ~~~1255 & ~~492 & ~~~~~446 & 11.38\% & 0.044\\
        \hline
    \end{tabular}
    \label{Tab:redshift}
\end{table*}

\begin{figure}
    \begin{tabular}{c}
	\includegraphics[width=\columnwidth, trim=1mm 1mm 1mm 2mm, clip]{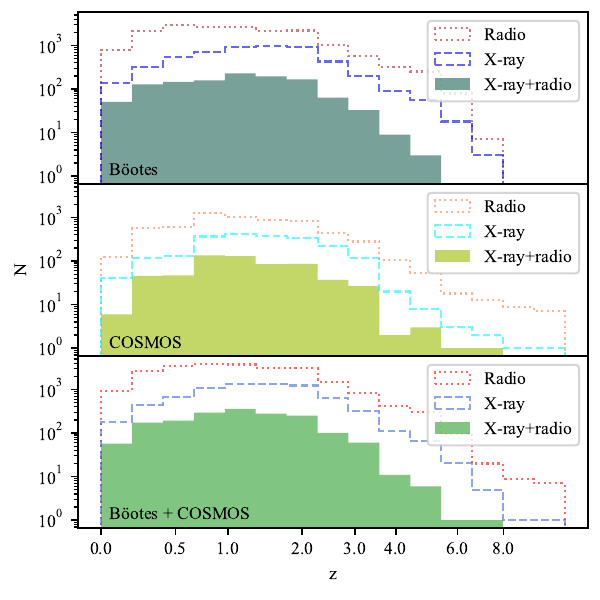}
	\end{tabular}
    \caption{Redshift distribution of the Bo\"otes and COSMOS survey samples used in this work, separated into radio and X-ray. They are a combination of spectroscopic (where available) and photometric redshifts. Of the sources with $z>0$, \tnum{6362/18,880} radio detected sources, \tnum{3878/7491} X-ray detected sources and \tnum{941/1538} detected X-ray+radio sources have spectroscopic redshifts.}
    \label{zhist}
\end{figure}

The distributions of redshifts for the radio and X-ray surveys in the Bo\"otes and COSMOS fields can be found in Figure \ref{zhist}, along with the distribution of the sources that are both X-ray and radio detected in both fields. The redshifts of this sample extend out to $z>8$ for both radio and X-ray detections, however, the vast majority of our cross-matched sample are below $z<4$. Our final cross-matched sample contains \tnum{1538} sources with both and X-ray and radio detection with a redshfit $z>0$, of which \tnum{941/1538} with spectroscopic redshifts.

\section{X-ray--radio luminosity comparison}
\label{sec:LXLRcomparison}

X-ray and radio detections are often both, individually, good tracers of AGN activity and previous work has shown that there is a correlation between the X-ray emission and the radio emission \citep[e.g. ][]{ 2003Merloni,Panessa2015,dAmato2022}. In this section, we directly compare the X-ray and radio luminosities of the X-ray and radio detected sample and compare to previously calculated correlations. 

Different telescopes/surveys have different frequency/energy ranges over which the flux densities of a source are measured. For a better comparison between the surveys used in this work, and with past literature, we calculate X-ray luminosities over the rest-frame 2--10~keV energy range and the radio luminosities at a rest-frame frequency of 1400 MHz. To do this we use a conversion factor. For the X-ray and radio, respectively, we use:
 \begin{equation}
     K_{\mathrm{X}} = \frac{(\rm{10~keV})^{2-\Gamma} - ({2~keV})^{2-\Gamma}}{{E_{\rm max,obs}}^{2-\Gamma} - {E_{\rm min,obs}}^{2-\Gamma}} 
 \end{equation}
 
 \begin{equation}
    K_{\mathrm{R}} = \Biggl(\frac{\rm1400~MHz}{\nu_{\rm obs}}\Biggr)^\alpha
\end{equation}
where $\Gamma(=1.9)$ is the photon index of the direct X-ray
emission and $E_{\rm min,obs}$ and $E_{\rm max,obs}$ are the minimum and maximum of the observed energy band range, respectively. For the radio, $\alpha=-0.7$ is the assumed spectral index of the radio emission and $\nu_{\rm obs}$ is the frequency the radio source was observed at.

The equations then used to convert between X-ray and radio observed flux densities and X-ray and radio luminosities, respectively, are:
\begin{equation}
    L_{\mathrm{X}} = \frac{4\pi D_{\rm L}^2f_{\mathrm{X}}K_{\mathrm{X}}}{(1+z)^{(2-\Gamma)}}
\end{equation}

\begin{equation}
    L_{\mathrm{R}} = \frac{4\pi D_{\rm L}^2f_{\mathrm{R}}K_{\mathrm{R}}}{(1+z)^{(1+\alpha)}}
\end{equation}
where $L_{\mathrm{X}}$ is the total luminosity over the rest-frame 2--10~keV band in units of erg/s, and $L_{\mathrm{R}}$ is the monochromatic luminosity per unit frequency evaluated at a rest-frame frequency of 1400~MHz, in units of W/Hz. The denominator of both equations is the k-correction for that wavelength, $D_{\rm L}$ is the luminosity distance and $f_{\mathrm{X}}$ and $f_{\mathrm{R}}$ are the observed X-ray (full/broad: 0.5--7~keV) fluxes and radio (COSMOS: 3~GHz; Bo\"otes: 144~MHz) flux densities, respectively.

\begin{figure}
\centering
    \begin{tabular}{c}
	\includegraphics[width=\columnwidth, trim=3mm 3mm 1mm 2mm, clip]{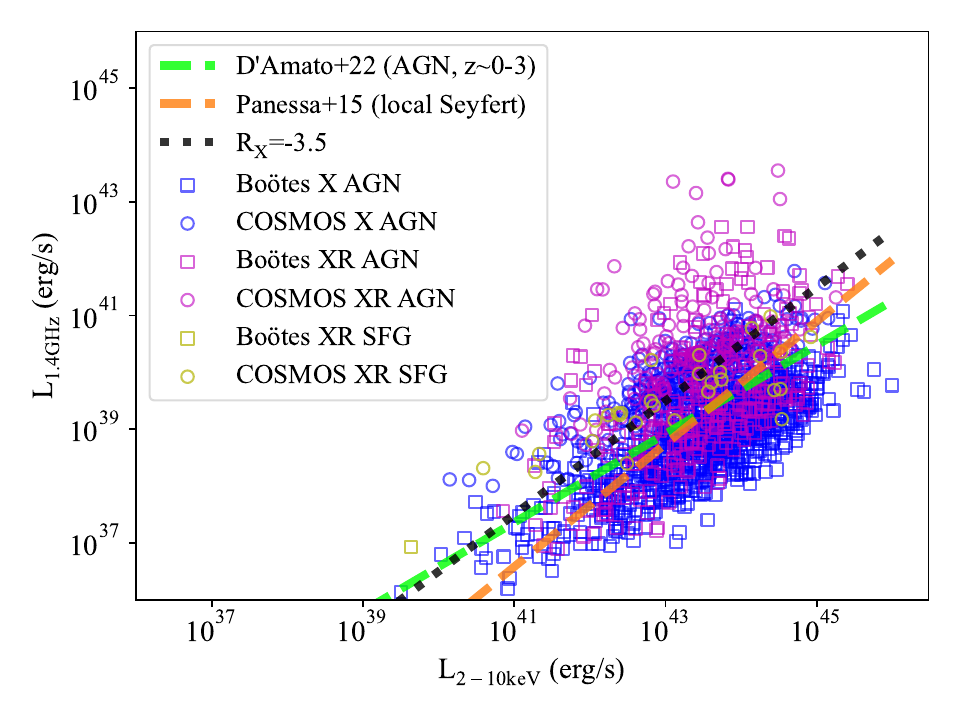}
	\end{tabular}
    \caption{Radio luminosity (1.4\,GHz) vs X-ray luminosity (2--10~keV) for the combined Bo\"otes and COSMOS sample, where a detection in both radio and X-ray is required to be displayed. Circles represent sources detected in the COSMOS field and squares represent those in the Bo\"otes field. Blue represents those that their X-ray is due to an AGN (`X AGN'), whilst magenta represents those whose radio and X-ray emission is due to an AGN (`XR AGN'). Yellow represents those where both the X-ray and radio emission is due to star-formation processes (`XR SFG'). There appears to be a good correlation between the two luminosities, albeit with a large scatter, especially towards higher $L_{\rm1.4~GHz}$. Over plotted are lines representing relations from \citet{dAmato2022} and \citet{Panessa2015}, as well as a black dashed line representing the divide between radio-loud and radio-quiet populations based on the radio loudness parameter, $R_{\rm X}$ = log($\frac{L_{\rm R}}{L_{\rm X}}$) \citep{Terashima2003,Lambrides2020}. Caution is, however, advised with using these relations; see Figure \ref{LxvsLr_zbins} for more information.}
    \label{LxvsLr}
\end{figure}

\begin{figure*}
\centering
    \begin{tabular}{c}
	\includegraphics[width=\textwidth, trim=1mm 0mm 1mm 2mm, clip]{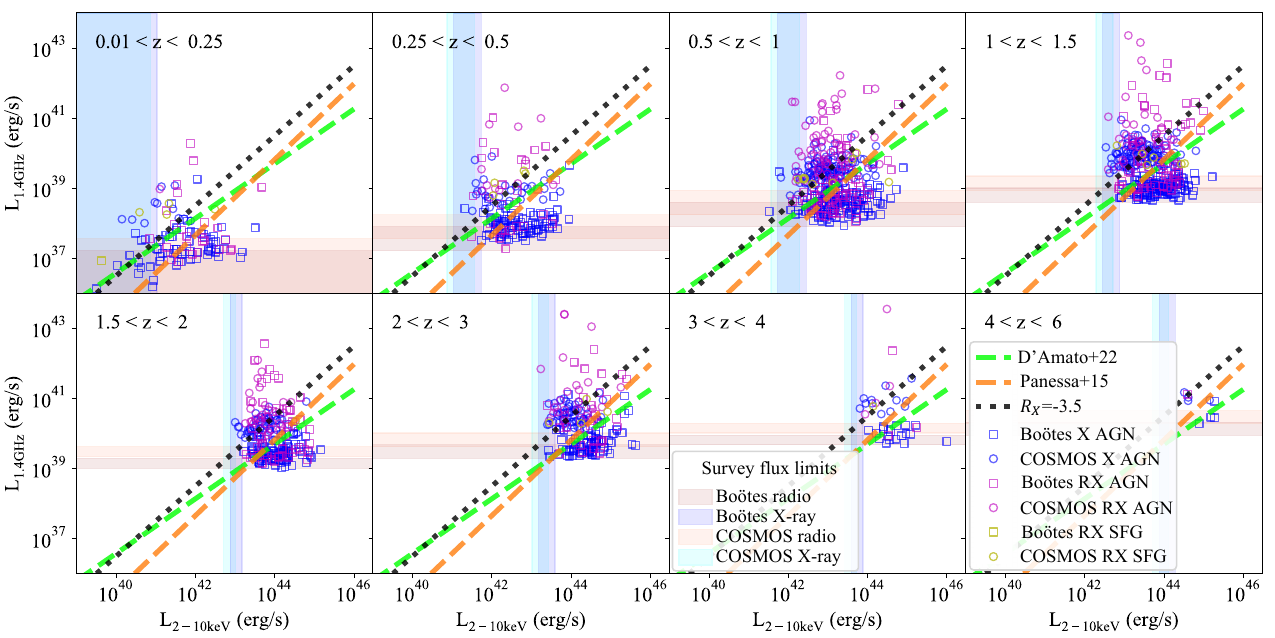}
	\end{tabular}
    \caption{Radio luminosity (1.4\,GHz) vs X-ray luminosity (2--10~keV) for the combined Bo\"otes and COSMOS sample, separated into redshift bins. Shaded regions represent the flux limits of the X-ray (blue) and radio (red) surveys converted to luminosity space from the beginning to the end of the redshift bins. Our results show the lack of a direct, underlying correlation between the X-ray and radio luminosities of AGN. These results show that the positive correlation, seen in Figure \ref{LxvsLr}, is primarily introduced due to the flux limits of the surveys used to identify them. As such, correlations between X-ray and radio luminosities presented in prior literature should be treated with caution. }
    \label{LxvsLr_zbins}
\end{figure*}

We combine the COSMOS and Bo\"otes radio and X-ray detected samples, numbering 1894 sources, 1544 of which have reliable redshifts and both radio and X-ray detections (COSMOS full band: $F_{\rm prob}<4\times10^{-6}$; Bo\"otes full band: $F_{\rm prob}<10^{-4.63}$, where $F_{\rm prob}$ is the Poissonian probability that the source is a background fluctuation). The luminosities at 1.4\,GHz and 2--10~keV were then calculated and plotted in Figure \ref{LxvsLr}. For the purpose of better comparison with $L_{\mathrm{X}}$ we convert the radio flux densities to luminosities in erg/s instead (we use $L_{\mathrm{R}}$ in units of W/Hz throughout the rest of this paper). From this we can see that the X-ray and radio luminosities show a good, positive correlation, although with large amounts of scatter\footnote{but see also, Figure \ref{LxvsLr_zbins}}.

The sources of the Bo\"otes surveys are separated into star-forming and AGN as described in \cite{Best2023} and \cite{2021Duncan} for the radio and X-ray, respectively. Star-forming and AGN dominant sources are separated as described in \cite{2017bSmolcic} and \cite{2016Marchesi} for the COSMOS field, for radio and X-ray, respectively. In the Bo\"otes field, for a source to be selected as an X-ray AGN the X-ray to optical flux ratio had to be $X/O = \log _{10}(f_{\mathrm{X}}/f_{\rm opt})>-1$, or the X-ray hardness ratio $>0.8$ \citep{2021Duncan}. In the COSMOS field, we took the estimated values of star-formation from the \citet{2016Marchesi} SED fitting and calculated the expected X-ray emission from star-formation \citep{Aird2017}, and those with X-ray emission in excess of a factor 2 above of this were labelled as X-ray AGN. For both the COSMOS and Bo\"otes fields, SED fitting, from UV to far-IR, was carried out for all of the radio detected sources, from which star-formation rates were estimated \citep{2017bSmolcic, Best2023}. Sources that had an excess of radio emission compared to expected estimates from star-formation were then classed as radio-excess AGN. In the case of the X-ray, in Figure \ref{LxvsLr} it can be seen that the sources with emission due to star-formation are few in number, as expected since an X-ray detection alongside a radio detection would most likely point to an AGN. Furthermore, the sources with both X-ray and radio emission due to an AGN have a tail towards higher radio luminosities, most likely representing the radio-loud population.

Over-plotted on Figure \ref{LxvsLr} are relations from previous literature that have calculated a relation between the radio and X-ray luminosity of AGN. The green dashed line represents \cite{dAmato2022}, which used a sample of 243 X-ray-selected objects (0 $<$ $z$ $<$ 3) obtained from 500 ks \textit{Chandra} observations of the J1030 equatorial field. Also plotted, in orange, is the X-ray/radio relation found for a sample of local ($z$ $<$ 0.35) X-ray-selected Seyfert galaxies by \cite{Panessa2015}. There is also a radio loudness parameter, represented by a black dotted line, where $R_{\rm X}$ = log($\frac{L_{\rm R}}{L_{\rm X}}$) $= -3.5$ is the divide between radio-loud and radio-quiet AGN, which was based on low-luminosity AGN samples \citep{Terashima2003,Lambrides2020}. These relations go through the middle of the data as expected, though there is $\sim$ 3 orders of magnitude of scatter around the relations.

The X-ray and radio detected sources were then separated into different redshift bins, as seen in Figure \ref{LxvsLr_zbins}, with the detection limits of the Bo\"otes and COSMOS surveys in X-ray and radio luminosity space also plotted. The positive correlation between X-ray and radio luminosity that we saw across all redshifts can no longer be seen, and there is no significant correlation between X-ray and radio luminosity at any given redshifts. Prior studies may be finding a correlation due to changing flux limits with redshift instead, as the relations shown by the green and orange dashed lines roughly line up with the cross-over of the flux limits in radio and X-ray luminosity space. Our results clearly show a very broad range of radio luminosities for a given X-ray luminosity (and vice versa), with sources stretching to the high radio luminosities, the radio loud regime, regardless of the X-ray luminosity.

\section{Measurements of the Luminosity Function}\label{sec:MeasLF}

Given the findings of Section \ref{sec:LXLRcomparison}, we seek to improve methods to accurately quantify the numbers of sources at different radio and X-ray luminosities, and the relationship (if any) between these quantities over cosmic time, that carefully accounts for the impact of the flux limits of the surveys used to identify the sources. 

In this section, we first outline our method (Section \ref{method}) for estimating the luminosity function in both X-ray and radio luminosity space and compare with measures that only consider flux limits in a single band. Then we apply this method to our X-ray and radio detected sample in Section \ref{sec:LF}. First, in Section \ref{sec:RLF} we apply the method to the overall radio sample at different redshifts, producing measurements of the overall radio luminosity function as well as for sub-samples at different ranges of X-ray luminosity. We do the same for the overall X-ray sample in Section \ref{sec:XLF}, producing measurements of the overall X-ray luminosity function and split into increasing ranges of radio luminosity. Next we determine the space densities of sources in X-ray and radio luminosity space at the same time, thus producing measurements of the multidimensional X--ray--radio luminosity function. Lastly, in Section \ref{sec:AGNdom}, we separate out the AGN dominated population and explore how the fraction of X-ray--radio detected AGN to the overall X-ray and radio detected AGN sample changes with increasing X-ray and radio luminosity.

\subsection{Method}\label{method}

The $\Sigma V_\mathrm{max}^{-1}$ estimator \citep{Vmax} has been widely used for  calculating binned luminosity functions, however it can suffer from biases when using small samples and when a large proportion of sources lie close to the survey flux limits \citep{PandC,Johnston2011}. We thus use and adapt the method proposed by \cite{PandC} to derive the LF as it should be less affected by systematic errors when adopting comparatively broad bins in luminosity and redshift that are impacted by the flux limit of the survey. \cite{NobsNmod} also proposed an alternative, referred to as the $N^{\rm obs}/N^{\rm mdl}$ estimator, that weights by the expected distribution of sources \emph{within} a given luminosity--redshift bin; however, this requires a known model, which we do not have for a combined X-ray--radio LF.

To estimate the luminosity functions for the individual X-ray and radio wavebands we use the method described in \cite{PandC}, modified to account for changing sensitivity across survey area:
\begin{equation}\label{PandCeq}
    \phi_{\rm est, \lambda} = \frac{N_{\lambda}(l_{\lambda},z)}{\int_{z_{\rm min}}^{z_{\rm max}} \int_{l_{\rm \lambda, min}}^{l_{\rm \lambda, max}}A(f_{\lambda}(L_{\lambda},z)) \frac{dV}{dz}dz \ d\log(L_{\lambda})}
\end{equation}

Where $N_\lambda$ is the total number of sources in a specified redshift and luminosity (X-ray or radio) range, $\lambda$ represents the wavelength at which the luminosity function is being measured (X-ray or radio) and $A$ is the survey area that is sensitive to a source with a specified flux (as defined by the sensitivity curves in the X-ray and the flux cuts in the radio), $f_\lambda$, corresponding to a given luminosity, $L_\lambda$, and redshift, $z$.
$\frac{dV}{dz}$ is the differential co-moving volume (with respect to $z$) per unit area. 
$l_{\rm \lambda,min},l_{\rm \lambda,max}$ (where $l_{\rm \lambda,min},l_{\rm \lambda,max}) \equiv (\log L_{\rm \lambda,min}, \log L_{\rm \lambda,max}$) and $z_{\rm min},z_{\rm max}$ indicate the limits of the luminosity and redshift bin over which the binned luminosity function is calculated.
1$\sigma$ equivalent uncertainties on
$\phi_{\rm est, \lambda}$ are calculated based on the Poisson uncertainties in the observed
source number, $N_\lambda$, using the relations given by \cite{1986Gehrels}.

\begin{figure*}
\centering
    \begin{tabular}{c}
	\includegraphics[width=0.9\textwidth, trim=0.5mm 1mm 1mm 0mm, clip]{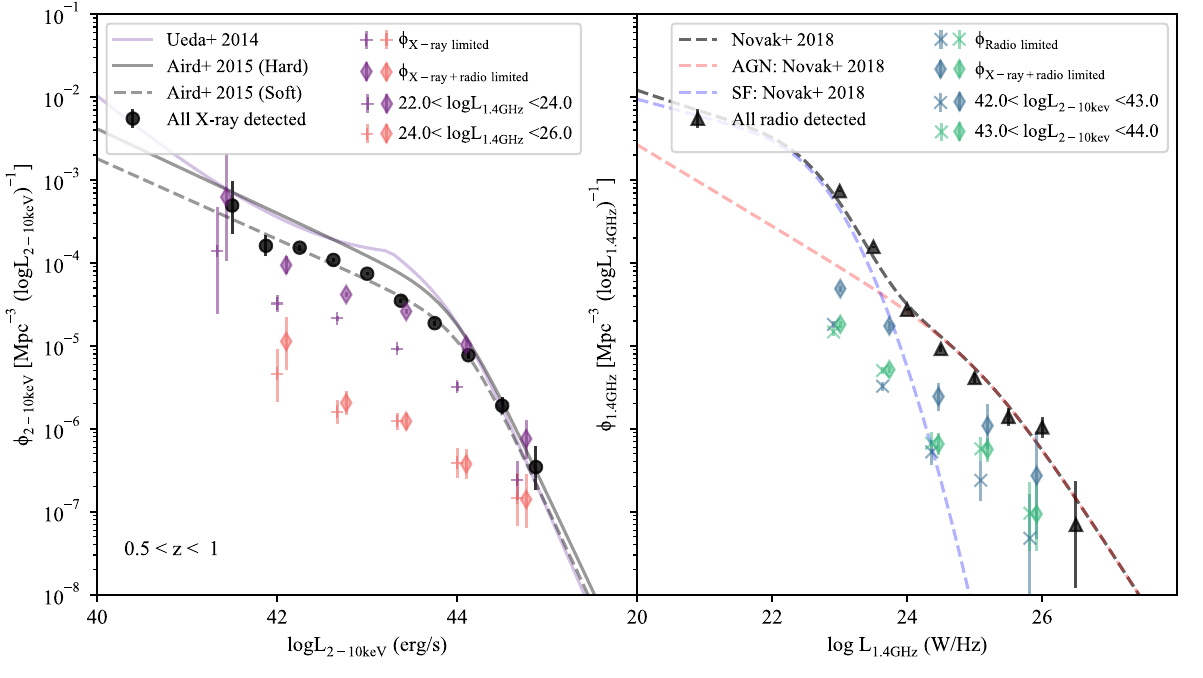}
	\end{tabular}
    \caption{Examples of the XLF (left) separated into different ranges of detected radio luminosities and RLF (right) separated into different ranges of detected X-ray luminosities for the combined Bo\"otes and COSMOS sample for 0.5 $<$ z $<$ 1. We illustrate the difference between using the standard Page \& Carrera method ($\rm\phi_{2-10~keV}$ and $\rm\phi_{1.4~GHz}$) that only accounts for limits in the corresponding waveband (`+' and `x' symbols in left and right panels, respectively) and using our updated method ($\rm\phi_{\mathrm{XR}}$), where the detection limits of both X-ray and radio have been taken into account (coloured diamonds in both panels). 
    In the left panel we plot XLF models from \citet{Ueda2014} (purple line) and \citet{Aird2015} (dashed grey and solid grey lines representing soft X-rays and hard X-rays, respectively) and in the right panel we plot the RLF model from \citet{Novak2018} (dashed lines), separated into AGN (red), star-forming (blue) and combined (black). All models are plotted at $z=0.75$.
    The updated method has a greater impact on the lower luminosity bins that are more prone to incompleteness for either wavelength.}
    \label{methodplot}
\end{figure*}

We use equation \ref{PandCeq} to estimate the XLF and RLF from the combined Bo\"otes and COSMOS sample (based on all sources detected in the corresponding wavelength, where the sensitivity curves for the X-ray have been added together to create a combined sensitivity curve). 
Figure \ref{methodplot} shows an example of such measurements for the $0.5<z<1.0$ redshift bin, where the black circles (left panel) shows measurements of the XLF based on the X-ray detected sample while the black triangles (right panel) shows measurements of the RLF based on the radio detected sample.
We also show estimates based on the subpopulation of the X-ray sample that is radio detected split into two different ranges of radio luminosities and vice versa (plus and cross symbols in Figure~\ref{methodplot} left and right, respectively). These measurements, however, only take into account the detection limits of either the X-ray or the radio band for the sources that are detected in both X-ray and radio, when for a source to enter our combined sample we should in fact take into account the detection limits of both the radio and X-ray data.

To estimate the X-ray--radio luminosity function we modify Equation \ref{PandCeq}, to calculate a single volume which allows for whether the radio or X-ray detection limits place the most stringent constraints on the survey area that we are sensitive to at a given $L_{\mathrm{X}}$ or $L_{\mathrm{R}}$ and $z$. By integrating over these joint limitations we are able to accurately assess the cosmological volume that we are probing with the combination of our X-ray and radio surveys for a given luminosity/redshift bin. 
Our binned estimator of the X-ray--radio luminosity function is thus given by

\begin{equation}\label{PandC3D}
    \phi_{\mathrm{XR}} = \frac{N_{\mathrm{XR}}(L_{\mathrm{X}},L_{\mathrm{R}}, z)}{\int_{z_{\rm min}}^{z_{\rm max}} \left[ \int_{l_{\rm X, min}}^{l_{\rm X, max}} \int_{l_{\rm R, min}}^{l_{\rm R, max}} A_{\rm m} d\log(L_{\mathrm{X}}) d\log(L_{\mathrm{R}}) \right] \frac{dV}{dz}dz }
\end{equation}

where $A_{\rm m} =min\Bigl(A\bigl(f_{\mathrm{X}}(L_{\mathrm{X}},z)\bigr), A\bigl(f_{\mathrm{R}}(L_{\mathrm{R}},z)\bigr) \Bigr)$ and $N_{\mathrm{XR}}$ is the total number of sources in a specified redshift, X-ray luminosity and radio luminosity range. 
While $\phi_{\rm est}$ has units of Mpc$^{-3}$ $(\log L_\lambda)^{-1}$ i.e. per unit volume per \emph{logarithmic luminosity interval}, $\phi_{\mathrm{XR}}$, which involves an integration over both the X-ray and radio luminosity interval, has units of Mpc$^{-3}$ ($\log L_{\mathrm{X}})^{-1} (\log L_{\mathrm{R}})^{-1}$. We note that to compare directly to measurements of luminosity functions in a single band thus requires us to multiply by the size of the bin in the other wavelength. We thus define $\phi_{\mathrm{X}}(L_{\mathrm{X}} | L_{\mathrm{R}})$, the X-ray luminosity function for sources of a given radio luminosity as
\begin{align}
    \phi_{\mathrm{X}}(L_{\mathrm{X}} | L_{\mathrm{R}}) &= \int_{l_{\rm R,min}}^{l_{\rm R,max}} \phi_{\mathrm{XR}}(L_{\mathrm{X}},L_{\mathrm{R}}) d \log (L_{\mathrm{R}}) \nonumber \\
    &\approx \phi_{\mathrm{XR}}(L_{\mathrm{X}},L_{\mathrm{R}}) \times \Delta \log L_{\mathrm{R}}
\end{align}
where $\Delta \log L_{\mathrm{R}} = l_{\rm R,max}-l_{\rm R,min}$ is the size of the logarithmic radio luminosity bin. 
The radio luminosity function of sources of a given X-ray luminosity, $\phi_{\mathrm{R}}(L_{\mathrm{R}} | L_{\mathrm{X}})$, is defined in an equivalent manner.

Figure \ref{methodplot} shows an example of how our method (diamond points) compares to using the original Equation \ref{PandCeq} (`+' and `x' points) that only considers the limitations due to the flux limit of a single band. We can see from this figure that using our new method increases our estimate of the LF, an effect which is more noticeable for the faintest X-ray luminosities in the RLF and the faintest radio luminosities in the XLF. 
The impact is greater at fainter luminosities as the sensitivity limits have the greatest effect and accounting for these in both bands correctly reduces our estimate of the volume over which sources could be detected, bringing the overall LF up.

\begin{figure}
\centering
    \begin{tabular}{c}
	\includegraphics[width=\columnwidth, trim=1mm 1mm 1mm 1mm, clip]{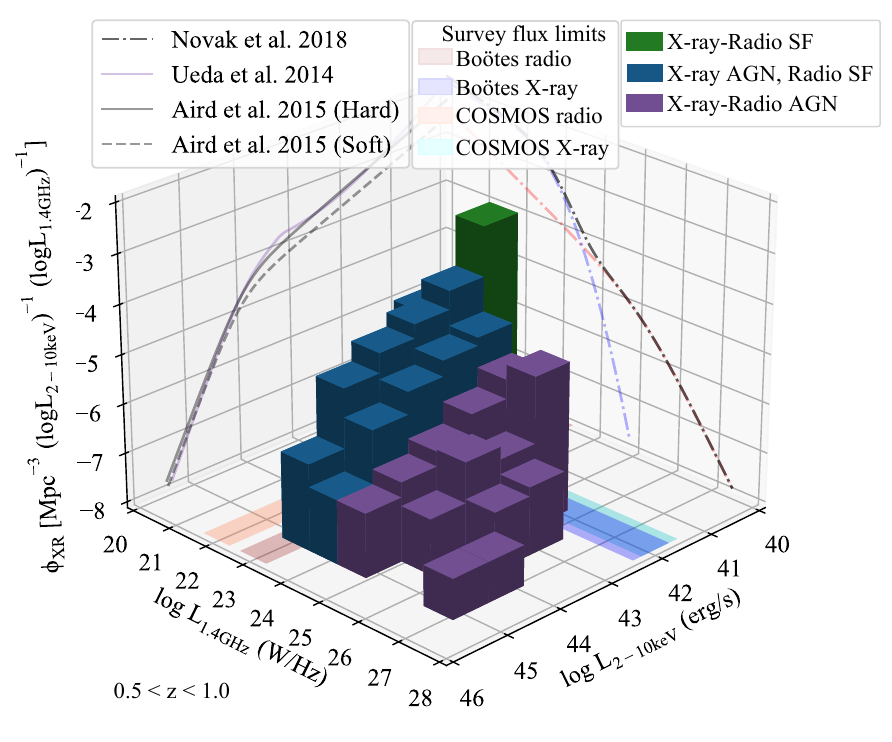}
	\end{tabular}
    \caption{X-ray radio luminosity function (XRLF), where the density of sources has been calculated per X-ray and radio luminosity bin. The different coloured bins represent the areas of the LF where either SF or AGN are expected to dominate in the radio and/or X-ray bands (see legend for details). It should be noted that in some redshift ranges we also show `Radio AGN, X-ray SF' in yellow, which can be seen in Appendix \ref{sec:XRLF}. The lines on the $\phi_{\mathrm{XR}}$ vs $\log L_{\rm 2 \text{--} 10~keV}$ axis are the XLF models from \citet{Ueda2014, Aird2015} and the lines on the $\phi_{\mathrm{XR}}$ vs $\log L_{\rm 1.4~GHz}$ axis is the RLF model from \citet{Novak2018}, separated into AGN and star-forming sources. All models are plotted at $z=0.75$. The shaded regions on the $\log L_{\rm 1.4~GHz}$ vs $\log L_{\rm 2 \text{--} 10~keV}$ axis represent the flux limits of the X-ray (blue) and radio (red) surveys converted to luminosity space from the beginning to the end of the redshift bins. See Appendix \ref{sec:XRLF} for equivalent 3D representations of the XRLF at different redshifts.}
    \label{example3D}
\end{figure}

Our new method provides an estimator of the LF that depends on \emph{both} X-ray and radio luminosity.
In Figure~\ref{example3D} we show a example 3D plot of these measurements, capturing the distribution of sources over a wide range in both luminosities. Here, we also distinguish star-forming dominant and AGN dominant populations based on the radio and X-ray luminosities. In X-ray we choose a boundary of $>$10$^{42}$ erg/s to define the AGN dominant population, and in radio we use a redshift-dependent cut based on models by \cite{Novak2018}, where the AGN dominant population is defined as when the AGN model density is greater than the star-forming model density ($\phi_{\rm AGN}>\phi_{\rm SF}$).
The 3D distribution clearly shows the rarity of sources that are \emph{both} X-ray and radio bright and the existence of a large population of sources where the X-ray emission is AGN dominated and spans the full $L_{\mathrm{X}}$ range but the radio emission is star-formation dominated (blue histogram in Figure~\ref{example3D}). It should be noted that while the blue histogram denotes where star-forming sources are expected to dominate over AGN sources in the radio, the sources here are also X-ray detected, which increases the likelihood these sources have radio emission due to AGN.

\subsection{Luminosity function measurements over the full redshift range}\label{sec:LF}

\subsubsection{The radio luminosity function}\label{sec:RLF}

\begin{figure*}
\centering
    \begin{tabular}{c}
	\includegraphics[width=\textwidth, trim=0.5mm 1mm 1mm 2mm, clip]{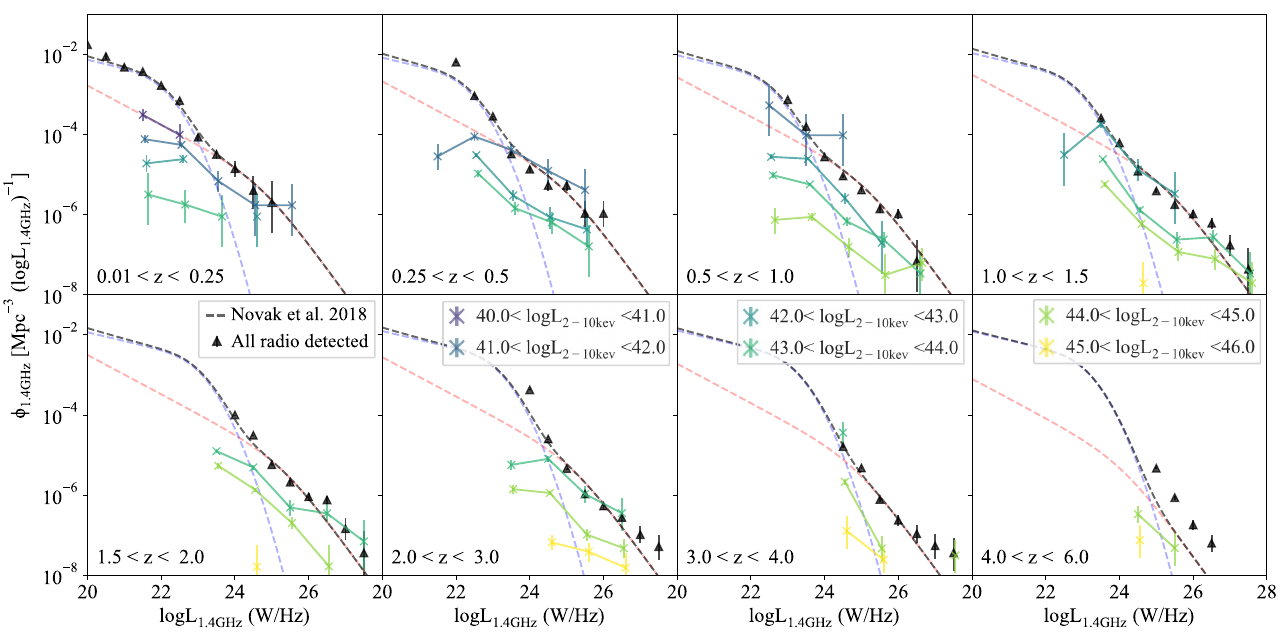}
	\end{tabular}
    \caption{RLF for the combined Bo\"otes and COSMOS sample for the full radio sample (black triangles), separated into different ranges of redshift and compared with the RLF model from \citet{Novak2018} (dashed lines), separated into AGN (red), star-forming (blue) and combined (black). All models are plotted at the mid-point of each redshift range. The RLF of the X-ray detected sample (coloured `x' symbols) is also plotted for different ranges of detected X-ray luminosities. This shows that for a given X-ray luminosity range, there is a broad range of radio luminosities.}
    \label{2DRXLF-RLF}
\end{figure*}

In Figure~\ref{2DRXLF-RLF}, we present measurements of the radio luminosity function using
the combined COSMOS and Bo\"otes radio-detected samples, separated into redshift bins (black triangles).
The RLF model from \cite{Novak2018}, which is separated into an AGN and star-formation component (red dashed and blue dashed lines, respectively, in Figure~\ref{2DRXLF-RLF}), shows good agreement with our RLFs including the sharp rise toward lower radio luminosities due to the increasing dominance of star-forming galaxies in the radio samples. We note that at the highest redshifts ($z>4$) the models appear to underestimate the RLF compared to our measurements. 

We also calculate the RLF of the X-ray detected sub-sample of radio sources, via Equation~\ref{PandC3D} and the methods described in Section~\ref{method} above to accurately account for the survey sensitivities at both X-ray and radio wavelengths. These measurements are shown for increasing ranges of X-ray luminosity by the coloured crosses in Figure~\ref{2DRXLF-RLF}. 
While, as expected, the most luminous X-ray sources have a lower space density i.e. are rarer, we find that the X-ray detected radio sources have a broad distribution of radio luminosities at all redshifts and for all $L_{\mathrm{X}}$ ranges.
As such, there does not appear to be a direct correlation between X-ray and radio luminosity, consistent with our findings in Section~\ref{sec:LXLRcomparison}. AGN of a given X-ray luminosity can produce a very wide range of radio luminosities. 
We note, however, that at the highest radio luminosities the X-ray limited measurements approach the RLF of the full radio sample, indicating the radio-brightest sources also \emph{tend} to be X-ray brighter, despite the lack of a direct correlation between $L_{\mathrm{X}}$ and $L_{\mathrm{R}}$.

\subsubsection{The X-ray luminosity function}\label{sec:XLF}

\begin{figure*}
\centering
    \begin{tabular}{c}
	\includegraphics[width=\textwidth, trim=0.5mm 1mm 1mm 2mm, clip]{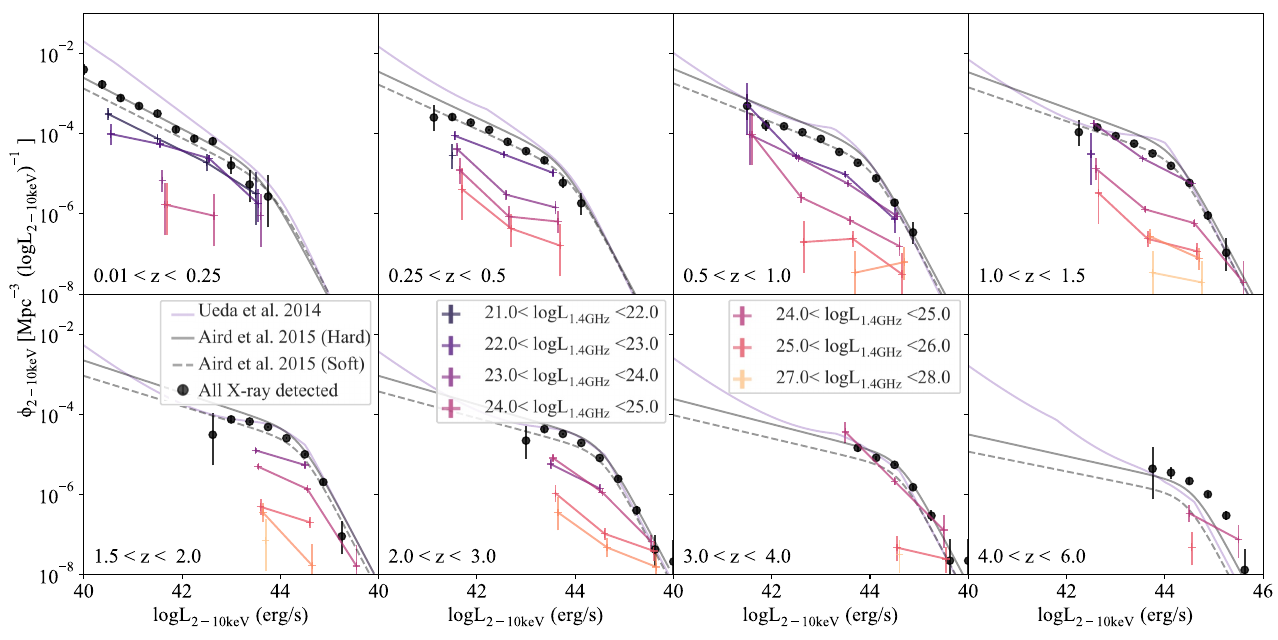}
	\end{tabular}
    \caption{XLF for the combined Bo\"otes and COSMOS sample for the full X-ray sample (black circles), separated into redshift bins and compared with the XLF models from \citet{Ueda2014} (purple line) and \citet{Aird2015} (dashed grey and solid grey lines representing soft X-rays and hard X-rays, respectively). All models are plotted at the mid-point of each redshift range. The XLF of the radio detected sample (coloured `+' symbols) is also plotted for different ranges of detected radio luminosities. This shows that for a given radio luminosity range, there is a broad range of X-ray luminosities.}
    \label{2DRXLF-XLF}
\end{figure*}

In Figure~\ref{2DRXLF-XLF}, we present measurements of the X-ray luminosity function using
the combined COSMOS and Bo\"otes X-ray samples, separated into redshift bins (black circles). 
We compare our measurements with parametric models from prior studies by \citet{Ueda2014} and \citet{Aird2015}. 
In general, our estimates agree well with these models, although we note that our measurements---based on fluxes measured in the full (0.5--7~keV) observed energy band---are closer to the models for the soft (0.5--2~keV) selected sample from \citet{Aird2015} at lower redshifts ($z\lesssim1.5$) but agree better with the hard (2--7~keV) model at higher redshifts. 
The discrepancy likely reflects the differing impact of intrinsic line-of-sight absorption (i.e. due to the AGN obscuring torus) on the observed energy bands at different redshifts.\footnote{
\citet{Aird2015} used the varying impact of absorption on the soft and hard X-ray bands at different redshifts to infer the underlying distribution of intrinsic line-of-sight equivalent hydrogen column densities ($N_\mathrm{H}$) and correct for these effects; in contrast, here we use observed fluxes in the full (0.5--7~keV) band to estimate luminosities, and neglect the impact of intrinsic $N_\mathrm{H}$ as a more advanced analysis is beyond the scope of this study.}
At the highest redshifts ($z>4$), all models lie below the observed data, as also seen for the RLF, suggesting that the space density of X-ray AGN may be higher than model extrapolations are indicating in the early Universe. This agrees with other work that has shown that the XLF models created for the low-$z$ regime do not work as well for the high-$z$ regime \citep[e.g.][Barlow-Hall et al. \textit{in prep.}]{Pouliasis2024}.

We also show measurements of the XLF of the radio detected X-ray sample for increasing ranges of radio luminosity (coloured `+' symbols in Figure~\ref{2DRXLF-XLF}), calculated using our updated method given by Equation~\ref{PandC3D} to account for both the X-ray and radio sensitivities. Similar to our findings for the X-ray detected radio sources (shown in Figure~\ref{2DRXLF-RLF} and discussed above), the radio detected X-ray sources shown in Figure~\ref{2DRXLF-XLF} have a broad distribution of X-ray luminosities for all radio luminosity ranges and at all redshifts. This finding again implies that there is no direct correlation between the X-ray and radio luminosities of AGN at any redshift.

\subsubsection{X-ray--radio luminosity function}\label{XRLFmeas}

\begin{figure*}
\centering
    \begin{tabular}{c}
	\includegraphics[width=\textwidth, trim=0.5mm 1mm 1mm 1.1mm, clip]{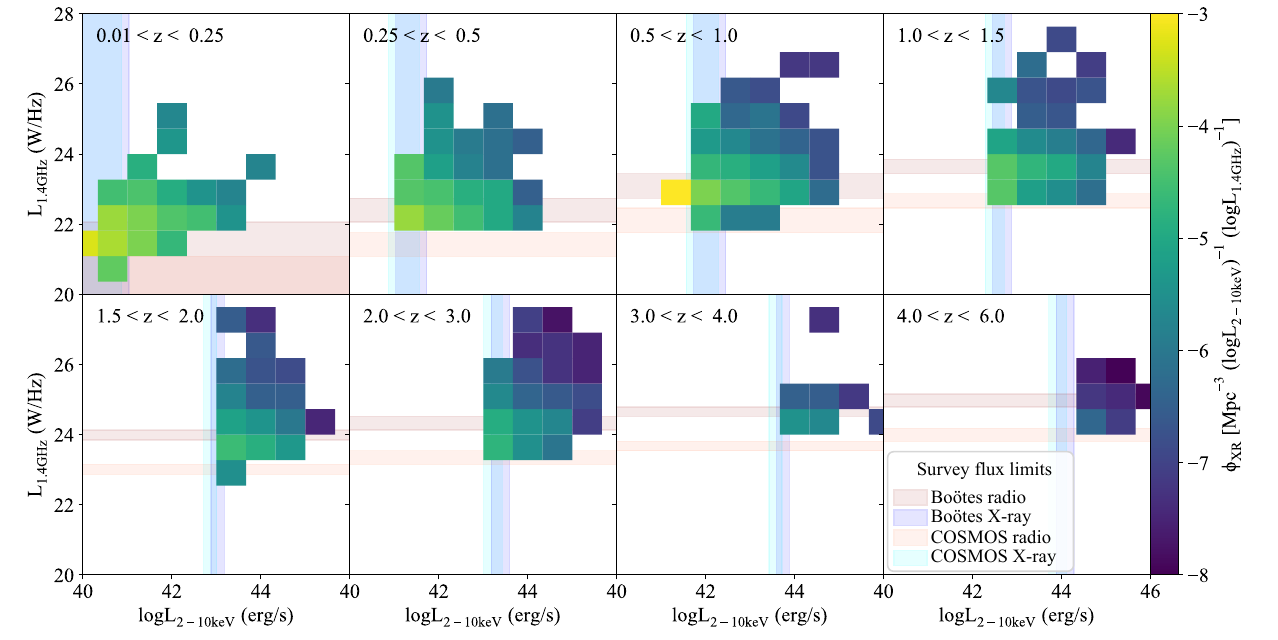}
	\end{tabular}
    \caption{Radio luminosity (1.4\,GHz) vs X-ray luminosity (2--10~keV) for the combined Bo\"otes and COSMOS sample, separated into redshift bins, with the colour representing the space density of sources over a corresponding radio and X-ray luminosity range. The coloured shaded regions indicate the X-ray or radio luminosities that correspond to the flux limits in the respective bands, with the width of the band indicating how the corresponding luminosity limits change over the redshift range. This shows that there is a broad, continuous distribution across both radio and X-ray luminosities. For 3D versions of these plots, see Appendix \ref{sec:XRLF}.}
    
    \label{FlatRXLF}
\end{figure*}

In Figure \ref{FlatRXLF} we present estimates of the X-ray--radio luminosity function (XRLF) in different redshift bins over the two-dimensional space of $L_{\mathrm{X}}$ and $L_{\mathrm{R}}$, where the colour indicates the space density of sources (i.e. the value of the XRLF).\footnote{3D representations of the XRLF measurements in each redshift bin are shown in Appendix~\ref{sec:XRLF}. A table of the measurements for each redshift bin will be made available alongside the paper.}
This plot shows that the XRLF is a broad, continuous distribution across both luminosities, limited by the detection limits of the surveys and the scarcity of the most luminous sources in a given area.

\subsection{AGN-dominated sample}\label{sec:AGNdom}

The measurements above quantify the broad distribution of radio luminosities of AGN of a given X-ray luminosity, and vice versa. Notably, we find that a large fraction of X-ray AGN hosts are producing radio luminosities that appear consistent with a star formation origin (below the point where star-forming galaxies dominate the space densities of the overall RLF), while a substantial fraction of the radio AGN hosts produce X-ray luminosities $\lesssim10^{42}$~erg~s$^{-1}$ that suggests a star formation origin.
In this section, we thus restrict our combined X-ray+radio sample to select sources in the AGN-dominated regime, based on the luminosity thresholds defined in Section \ref{method}.

\begin{figure*}
\centering
    \begin{tabular}{c}
	\includegraphics[width=\textwidth, trim=0.5mm 1mm 1mm 1mm, clip]{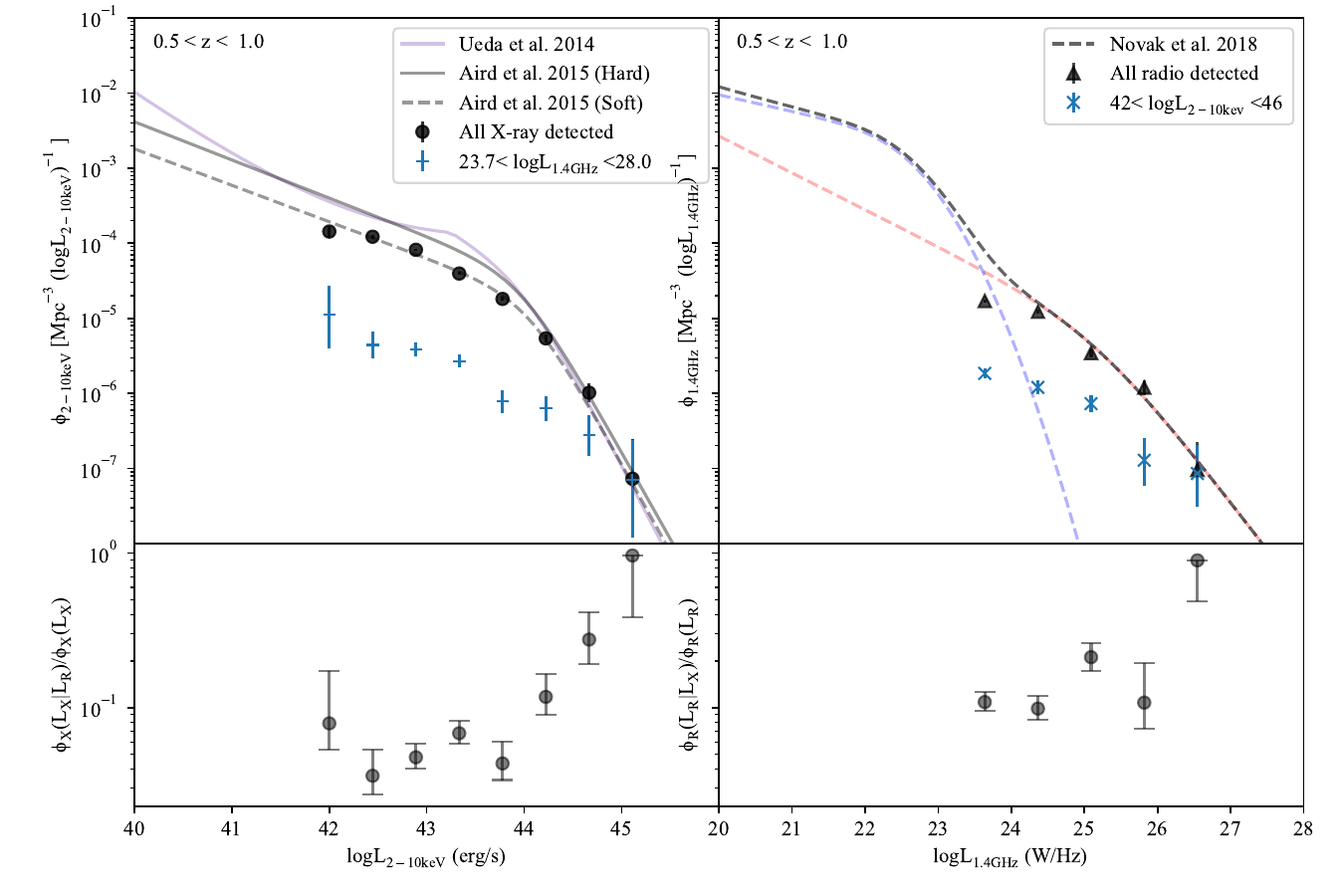}
	\end{tabular}
    \caption{X-ray and radio luminosity limited sample, where the majority of non-AGN sources have been removed. Shows the XLF (upper left) and RLF (upper right) of this sample, as well as the ratio of the space densities of the radio detected X-ray sources to the overall population of X-ray detected sources (lower left) and the ratio of the space densities of the X-ray detected radio sources to the overall population of radio detected sources (lower right). For the upper panels, the lines and points are as in Figures \ref{2DRXLF-RLF} \& \ref{2DRXLF-XLF}.}
    \label{RXLF_ratio_example}
\end{figure*}

In Figure \ref{RXLF_ratio_example} we plot the XLF and RLF of the X-ray+radio selected AGN that satisfy both thresholds and compare to the XLF and RLF of the overall X-ray and radio populations. 
These measurements allow us to isolate the fraction of the X-ray source population that are ``bona fide'' \emph{radio} AGN, and the fraction of the radio source population that are ``bona fide'' \emph{X-ray} AGN. 
In both cases, at low-to-moderate luminosities the X-ray+radio measurements (blue `+' and `$\times$', respectively) lie significantly below the single-band measurements (black circles and triangles, respectively) in Figure \ref{RXLF_ratio_example}, showing that a substantial fraction of the X-ray AGN population would not be identified as a radio AGN and that a substantial fraction of the radio AGN would not be identified as X-ray AGN \citep[which agrees with the results found in][]{Radcliffe2021}. 

To quantify these fractions more directly, we calculate the ratio of the XLF of the X-ray+radio AGN sample to the overall XLF,
\begin{equation}
\frac{\phi _\mathrm{2-10~keV}(L_\mathrm{2-10~keV}|L_\mathrm{1.4~GHz})}{\phi _{\rm 2-10~keV}(L_\mathrm{2-10~keV})}= \frac{\phi _{\rm X}(L_{\mathrm{X}}|L_{\mathrm{R}})}{\phi _{\rm X}(L_{\mathrm{X}})}
\end{equation}
and the ratio of X-ray+radio RLF to the overall RLF 
\begin{equation}
\frac{\phi _{\rm 1.4~GHz}(L_{\rm 1.4~GHz}|L_{\rm 2-10~keV})}{\phi _{\rm 1.4~GHz}(L_{\rm 1.4~GHz})} = \frac{\phi _{\rm R}(L_{\mathrm{R}}|L_{\mathrm{X}})}{\phi _{\rm R}(L_{\mathrm{R}})}.
\end{equation}
Errors on these fractions are calculated based on the binomial distribution, as in \cite{Cameron2011}, to capture the uncertainties due to the limited sample sizes used to measure both numerator and denominator. 
These ratios are shown in the lower panels of Figure~\ref{RXLF_ratio_example} for the $0.5<z<1.0$ range.
We find that $\sim5$\% of low-to-moderate X-ray luminosity AGN at these redshifts are also selected as radio AGN (lower left panel of Figure~\ref{RXLF_ratio_example}); however, the fraction rises at $L_{\mathrm{X}} > 10^{44}$~erg~s$^{-1}$, reaching 100\% in the highest X-ray luminosity bin.
Higher X-ray luminosity AGN are thus more likely to also be radio-selected AGN, though may not necessarily produce a high radio luminosity.
We also find that $\sim10$\% of moderate-luminosity radio AGN are also X-ray selected at these redshifts, with the fraction possibly rising toward higher radio luminosities (bottom right panel of Figure~\ref{RXLF_ratio_example}).

\begin{figure*}
\centering
    \begin{tabular}{c}
	\includegraphics[width=0.9\textwidth, trim=0.5mm 1mm 1mm 0mm, clip]{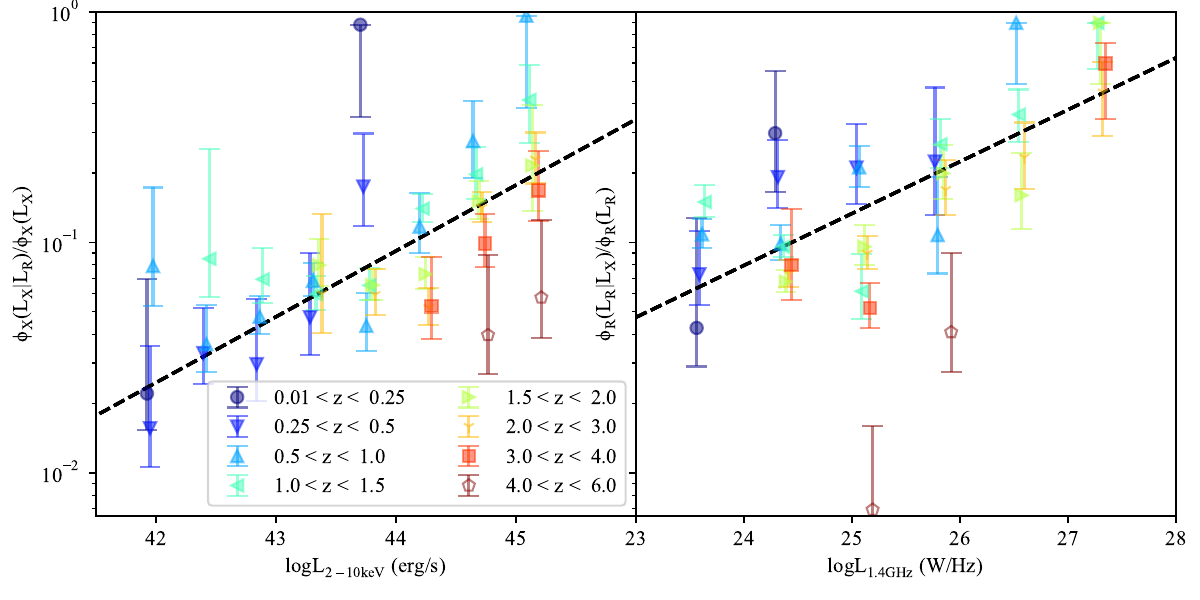}
	\end{tabular}
    \caption{(Left) The ratio of space densities of the radio detected X-ray sources to the overall X-ray detected sources. (Right) The ratio of space densities of the X-ray detected radio sources to the overall radio detected sources. The black dashed line is the $\chi^2$ best fit, but is not statistically a good fit. The $4.0<z<6.0$ bin is represented by an unfilled pentagon to represent that it is an incomplete bin and is not used in the fit. } 
    \label{RXLF_ratio_only}    
\end{figure*}

We then repeat the calculation of this ratio for each redshift bin, which can be seen in Figure \ref{RXLF_ratio_only}. Note that where a luminosity bin contained $\leq 2$ sources, the ratio was not plotted. Our measurements show that towards higher luminosities (in X-ray and radio) that an AGN is more likely to be both radio and X-ray selected, though will not necessarily be bright in both bands.

These measurements reproduce and carefully quantify established knowledge regarding the AGN population: that not all radiatively efficient (i.e. X-ray emitting) AGN produce strong radio emission; and that a substantial proportion of the radio AGN population correspond to a radiatively inefficient, jet-dominated accretion mode \citep[e.g. ][]{Heckman2014} that does not produce the typical AGN signatures at other wavelengths (e.g. X-ray emission). 
In addition, assuming that radio AGN that are also X-ray selected correspond to a radiatively efficient source, then our measurements imply that the brighter in radio luminosity that a source is, the more likely it is to be radiatively efficient. Such a pattern is in agreement with \cite{2025Kondapally}, who found that radio-detected AGN that were radiatively efficient were more likely to have an X-ray detection than radiatively inefficient radio-detected AGN.

A $\chi^2$ fit was used to fit a power law function to the ratio as a function of $L_{\mathrm{X}}$ and $L_{\mathrm{R}}$ across all redshifts, which can be seen as the dashed lines in Figure \ref{RXLF_ratio_only}. Note that the redshift range $4<z<6$ was not included in the fits due to the increased incompleteness in this bin, caused by the X-ray and radio surveys not reaching the required depth. The $0.01<z<0.25$ was also not included due to this range being volume limited, causing the number of sources in this bin to be reduced compared to other bins. These best-fitting relations were then statistically compared with all of the measurements and those in individual redshift bins to determine the goodness of fit. We first determined the overall $\chi^2$ for all measurements and in each redshift bin. Then, we took the p-value given by the $\chi^2$ distribution, that gives the probability of obtaining the measured $\chi^2$, or larger, given the size of the errors and the number of degrees of freedom (given by the number of data points) if the power law relation is correct. We measure p-values$<$0.01 in all but 2 redshift bins, indicating the relations are not good fits. The highest redshift bin, 4 $< z <$ 6, is the furthest from the rest of the data points, which is not unexpected, as this bin has the fewest sources from which to calculate these ratios and the majority have a photometric redshift (only 3/14 sources have a spectroscopic redshift).

\begin{figure*}
\centering
    \begin{tabular}{c}
	\includegraphics[width=0.9\textwidth, trim=0.5mm 1mm 1mm 0mm, clip]{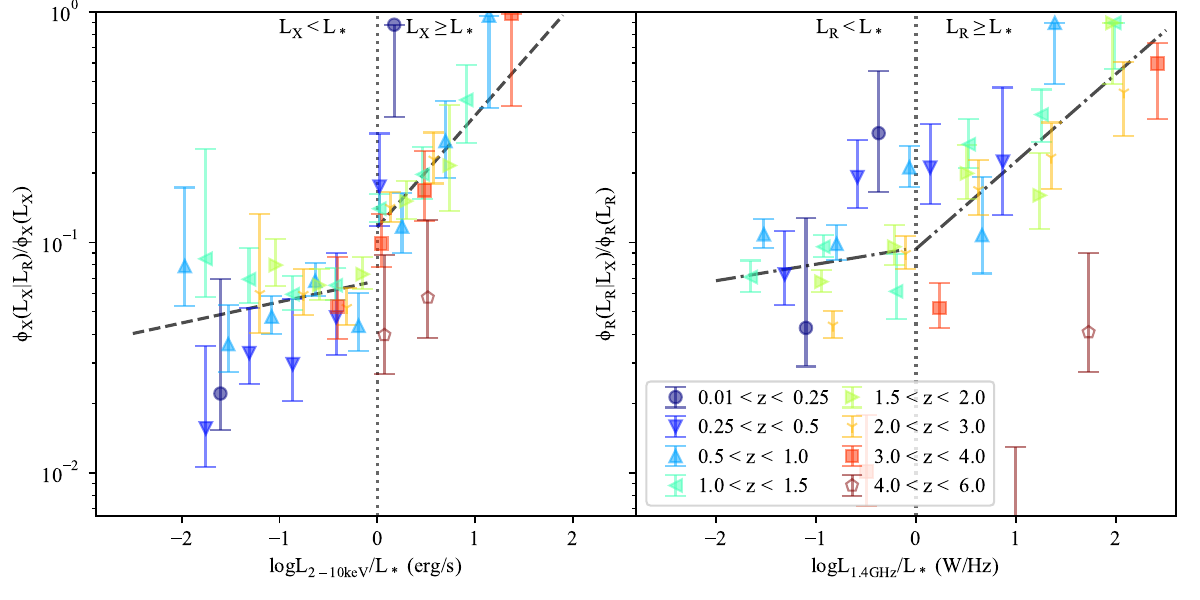}
	\end{tabular}
    \caption{(Left) The ratio of space densities of the radio detected X-ray sources to the overall X-ray detected sources with respect to $L_{\rm X}/L_*$. (Right) The ratio of space densities of the X-ray detected radio sources to the overall radio detected sources with respect to $L_{\rm R}$$/L_*$. The vertical dotted black line represents where $L_{\rm X}$$=L_*$ and $L_{\rm R}$$=L_*$, where $L_*$ is from the LF models from \citet{Aird2015} and \citet{Novak2018}, respectively. The black dashed line is the $\chi^2$ best fit, separated into $L<L_*$ and $L>=L_*$. The best fit lines for $L_{\mathrm{X}}\geq L_*$ (left) and $L_{\mathrm{R}}$$\geq$$L_{\mathrm{*}}$ (right) have p-values$\geq$0.01, showing good fits. The best fit lines for $L_{\mathrm{X}}$$\leq$$L_{\mathrm{*}}$ (left) and $L_{\mathrm{R}}$$\leq$$L_{\mathrm{*}}$ (right) have p-values$<$0.01, showing poor fits. The $4.0<z<6.0$ bin is represented by an unfilled pentagon to represent that it is an incomplete bin and is not used in the fit.} 
    \label{RXLF_ratio_Lstar}    
\end{figure*}

By comparing the overall RLF and XLF and luminosity functions of the X-ray--radio sample, e.g. Figure \ref{RXLF_ratio_example}, it can be seen that below the characteristic break in the XLF (i.e. at $L_{\mathrm{X}}\lesssim L_*$) that $\phi _{\rm X}(L_{\mathrm{X}}|L_{\mathrm{R}})/\phi _{\rm X}(L_{\mathrm{X}})$ is predominantly flat, whereas after the turnover at $\sim L_*$ that $\phi _{\rm X}(L_{\mathrm{X}}|L_{\mathrm{R}})/\phi _{\rm X}(L_{\mathrm{X}})$ increases. A similar trend, though not as clear, can be seen for the RLF. Therefore, instead of plotting the ratios of space densities against luminosity, we plot against $L/L_*$, which is shown in Figure \ref{RXLF_ratio_Lstar}. $L_*$ is taken from the LF models of \cite{Aird2015} and \cite{Novak2018} for the X-ray and radio, respectively.

For the ratio of space densities against $L/L_*$ we fit a power law to the data before and after the $L_*$ value in X-ray and radio luminosity space. For the XLF, at $L_{\mathrm{X}}\ge L_*$, we find that the ratio of space densities shows a good correlation with $L_{\rm X}$$/L_*$, which is independent of redshift. We see the same for the RLF but the statistical evidence for a correlation is weaker. The fits below $L_{\mathrm{X}}<L_*$ and $L_{\mathrm{R}}<L_*$ both give low p-values indicating a poor fit and that a single relation between the ratio and $L/L_*$ across all redshifts may be an over-simplification.

It should be noted that for $z>1.5$ the X-ray+radio AGN selected sample has increasing incompleteness of the X-ray AGN selection ($42<\log L_{2-10~\rm keV}<46$), whereas the radio AGN selection does not suffer from the same effects, most likely due to its redshift dependence. Therefore, the X-ray detected fraction is incomplete at higher redshifts, whereas the radio detected fraction of the X-ray sample is more complete. This could be a possible explanation to why the function fit of $\phi_{\mathrm{XR}}/\phi_{\mathrm{X}}$ at $L_{\mathrm{X}}$$>$$L_{\mathrm{*}}$ is a better fit than for $\phi_{\mathrm{XR}}/\phi_{\mathrm{R}}$ at $L_{\mathrm{R}}$$>$$L_{\mathrm{*}}$.


\section{Discussion}\label{sec:disc}

\subsection{X-ray--radio correlation (or lack thereof)}

The "Fundamental Plane of Black Hole Activity" \citep[e.g. ][]{2003Merloni, Falcke2004} suggests a clear correlation between black hole accretion and a relativistic jet, where X-ray emission is used as proxy for accretion rate and radio emission is used as a measure of the AGN jet. However, this tends to only apply to strong radio-loud AGN, as weaker radio sources deviate from the relation by several orders of magnitude \citep[e.g.][]{2008Li, 2022Bariuan}. Deviations from this correlation have been attributed to differences in the efficiencies of the accretion mechanisms \citep[e.g.][]{2003Gallo, 2003Merloni,dAmato2022}, where steeper slopes indicate an increase in accretion efficiency \citep[e.g. ][]{Coriat2011, Dong2014}, indicating sources accreting at higher Eddington ratios. In this work we are investigating the scatter in the luminosities around these expected correlations, in a sample with a broad range of X-ray and radio properties, across a broad redshift space.  

Previous studies of the relation between X-ray and radio luminosities tend not to include the sensitivity limits of the surveys, except to include an upper limit on those sources undetected in one band or the other. In Figure \ref{LxvsLr_zbins}, we show when split into redshift bins that, as expected, higher redshift sources in our sample are typically more luminous in both X-ray and radio, although their luminosities are scattered above the limits of the surveys in both bands. Furthermore, previously reported correlations appear to follow these flux limits, implying that they are tracing the limits of the detectors used to observe these sources. \cite{Siebert1996} suggested that, though they find a strong correlation between the soft X-rays and the core radio luminosity, the correlation of $L_{\mathrm{X}}$ with total $L_{\mathrm{R}}$ is most likely caused by redshift dependence of both X-ray and radio luminosities and/or the strong correlation of total $L_{\mathrm{R}}$ with its core $L_{\mathrm{R}}$. Furthermore, \citet{Mingo2014} propose that the positive correlations between $L_{\mathrm{X}}$ and $L_{\mathrm{R}}$ are the result of selection bias (such as a lower and upper limit on $L_{\rm Edd}$ of the sources in the sample).

Past literature exploring the correlation between X-ray and radio luminosities for the broader population of AGN has had a tendency to take either a local sample \citep[e.g. $z<0.05$;][]{2003Merloni, Magno2025} or combine a sample spanning a broad redshift range \citep[e.g. $0<z<3$; ][]{dAmato2022} that is used to fit a relation, without comparing different redshift ranges. Without careful consideration of the impact of flux limits, this can introduce a strong \emph{apparent} correlation between luminosities that is mainly driven by the flux limit and how it interacts with redshift (see e.g. our Figures \ref{LxvsLr} and \ref{LxvsLr_zbins}). What results is a relation that suggests a much tighter relationship between the X-ray emission (primarily tracing the accretion rate through a radiatively efficient disk) and radio emission (primarily coming from a jet) over cosmic time, when our measurements suggest a very high degree of scatter and that the instantaneous accretion rate is a somewhat poor predictor of the overall jet power, and vice versa. \citet{LaFranca2010} (sample of 375 X-ray selected AGN with radio detections, limited to log$L_{\mathrm{X}} \geq42$) produce a relation by first fitting a probability distribution function of $R_{\mathrm{X}} = \log$($L_{\mathrm{R}}/$$L_{\mathrm{X}}$) as a function of $L_{\mathrm{X}}$ and redshift using $\chi ^2$ estimators. It is based on the comparison of the observed and expected numbers of AGNs in $L_{\mathrm{X}}$-$z$-$R_{\mathrm{X}}$ space, obtained by taking
into account observational selection effects, such as radio flux limits, for each sample. They found that this method produces flatter relations than found by prior studies. We explore this more in Section \ref{sec:discLF} in the context of our XRLF measurements.

Soft X-rays have been found to correlate with high frequency radio flux, which has been proposed to be related to jet emission in radio-loud sources \citep[e.g.][]{Mingo2014}. However, large-scale jetted sources only account for $\sim$10\% of the AGN population \citep[e.g.][]{Best2005}, but for the less luminous, more numerous sources, those often referred to as radio-quiet, the origin of any radio emission is less clear and has been attributed to an unresolved, compact, sub-kpc nuclear region \citep[e.g.][]{Maini2016, Herrera2016}. Various theories for the origin of this compact radio emission have been proposed, such as a smaller-scaled version of relativistic jets, an AGN-driven wind, coronal emission, AGN photon-ionization of ambient gas (free-free emission), or star formation \citep[see ][and references therein]{Panessa2019}. It has been argued that a correlation between X-ray and radio luminosities for radio-quiet sources would indicate the radio is being produced from the same region the X-ray emission is produced from, i.e. the corona \citep[e.g. ][]{Laor2008}. It is also possible that it is not only one mechanism that produces this lower radio luminosity AGN, but several at once.

In some sources the X-ray corona may be the base of the radio jet \citep[e.g.][]{2015Wilkins}, which suggests a physical origin for any correlation. The radio frequencies expected to be detected from the base of the jet are between 10--100\,GHz, from which we would expect a tight correlation with the X-ray if the X-ray emission is indeed being produced by the same structure, however these frequencies are significantly higher than the frequencies used in this work. Detections of the base of the jet below $<10$~GHz are not expected due to strong synchrotron self-absorption in the dense vicinity of the corona \citep[e.g.][]{Inoue2014, Behar2018}. \citet{Panessa2007} found correlations between X-ray emission and radio emission at observed frequencies of 1.4, 5 and 15~GHz to have similar levels of significance and \citet{Panessa2015} found that X-ray correlates better with overall radio emission than core radio emission, which would not support the theory of the base of the jet being the X-ray corona. However, \citet{Baldi2022} found that X-ray emission correlated more tightly with frequencies of 45~GHz than 5~GHz, which would support the theory.  

X-ray and radio emission work on vastly different timescales and the X-ray and radio observations used in this work were not taken simultaneously. X-ray luminosity is a relatively instantaneous measurement, whereas radio luminosity is estimated from the weighted average over the whole radio lifetime of the source, contaminated by diffuse (older) emission from components such as lobes on a scale of a few pc to kpc which could take thousands of years to form. Even when simultaneous data has been used, only weak, though steep, correlations are found \citep[e.g. ][]{King2011, King2013, Jones2011}. 

The scatter around the correlations in the ($L_{\mathrm{X}}$, $L_{\mathrm{R}}$) plane have been suggested to be due to several intrinsic effects of the sources and how they are observed. For example, orientation, beaming, variability on different timescales and environmental interference \citep{Mingo2014}. Doppler beaming alone has been estimated to be able to introduce up to 3 dex of scatter \citep[e.g. ][]{Hardcastle1999}. Our results suggest no direct correlation, however, any underlying correlation could be blurred out by all these effects (timescales, Doppler beaming, etc.). Ultimately, the power in a jet may be fundamentally linked to the accretion power, and thus the X-ray emission, but our observations are not able to reveal such a complex connection.

\subsection{Quantifying the X-ray--radio connection through luminosity functions}\label{sec:discLF}

Our measurements of the XRLF provide a detailed quantification of the space densities of AGN with different X-ray and radio luminosities, and the relation between these quantities, over cosmic time.
Whilst, to our knowledge, an XRLF has not been measured previously, attempts have been made to use the relations between X-ray and radio luminosities to convert from an XLF to an RLF. For example, \citet{Mazzolari2024}, focusing on the COSMOS sample, use the redshift-independent relation found by \citet{dAmato2022} to estimate the RLF from their model of the hard XLF. These predicted RLFs were then compared to the RLF at different redshifts measured by \citet{2017bSmolcic}. They found that their model predicted higher space densities than observed at lower $L_{\mathrm{R}}$, which they attribute to the \citet{2017bSmolcic} measurements only being those with radio excess over expected radio emission from star-formation (based on the radio--FIR relation), and thus fainter radio AGN are missed as they are overwhelmed by the emission of the host galaxy. Their model also produces lower space densities than observed by \citet{2017bSmolcic} at higher $L_{\mathrm{R}}$, which they attribute to the \citet{dAmato2022} relation not taking into account the radio-loud population. 
Our work, however, shows that there is a broad range of possible radio luminosities for any given X-ray luminosity across all redshifts, thus converting an XLF into a RLF is not as straightforward as assumed in the \citet{Mazzolari2024} model.

\cite{LaFranca2010} studied the radio emission of hard X-ray-selected AGN and fit a probability distribution function of the ratio between the radio and X-ray luminosity, $R_{\mathrm{X}}$ = log(L$_{\rm 1.4~GHz}$/L$_{\rm 2-10~keV}$), for AGN of a given X-ray luminosity and redshift, P($R_{\mathrm{X}}$ | $L_{\mathrm{X}}$, $z$). 
To compare to their results, we determine that 
\begin{equation}
P(R_{\mathrm{X}} | L_{\mathrm{X}}, z) \equiv \frac{\phi_{\rm XR}(L_{\mathrm{R}} | L_{\mathrm{X}}, z)}{\phi_{\rm X}(L_{\mathrm{X}}, z)}
\end{equation}
evaluated at a given $L_{\mathrm{X}}$ 
and can thus be derived from our measurements of the XRLF and XLF.
$L_{\mathrm{X}}$ and $z$ are taken as the mid-point of the given X-ray luminosity and redshift bins, respectively. Our estimates of P($R_{\mathrm{X}}$ | $L_{\mathrm{X}}$, $z$) for different $z$ and $L_{\mathrm{X}}$ bins can be seen in Appendix \ref{Lafrancasec}, Figure \ref{Lafrancaplot}. We find that the model of \cite{LaFranca2010}, which was based on just 375 X-ray and radio detected sources, agrees well with the values calculated through our XRLF method, based on  a sample of \tnum{1538} X-ray and radio detected sources. Deviations from the model are found at the limits in ($L_{\mathrm{X}}$, $z$) space for our sample, and could be due to incompleteness in those ($L_{\mathrm{X}}$, $z$) bins, in the analysis done in \citet{LaFranca2010}.

The probability distribution shows that for a given $L_{\mathrm{X}}$ and $z$ value, there is a peak in the probability distribution of $L_{\mathrm{R}}$.  The location of this peak with regards to $L_{\mathrm{R}}$ shifts towards lower $L_{\mathrm{R}}$ with increasing $L_{\mathrm{X}}$, and shifts towards higher $L_{\mathrm{R}}$ with increasing $z$. The increasing of redshift shifting the peak towards higher $L_{\mathrm{R}}$ could be due to selection effects (only the brightest radio sources are detected towards higher redshifts). Our sample follows both the trend with redshift and $L_{\mathrm{X}}$. 
As we extend towards lower $L_{\mathrm{R}}$ it becomes harder to disentangle the SF and AGN radio emission. 

\cite{LaFranca2010} used their probability distribution fit, restricted to sources with log$L_{\mathrm{X}}$>42, to convert the XLF into an RLF and then compared with the measured RLF from \citet{Smolcic2009}. This provided a fairly good reproduction of the FRI (radio core-bright sources) RLF, though shows source densities slightly higher than measured at low $L_{\mathrm{R}}$ for $z$ = 0.2--1, which they suggest is because of radio emission due to star formation in the AGN hosting galaxies, which potentially affected their measure of P($R_{\mathrm{X}}$) at the lowest values of $R_{\mathrm{X}}$.  
\citet{LaFranca2010} and \citet{Mazzolari2024} both show that they can model the expected source counts well when they take into account a missing subsample (FR II and radio-loud, respectively), though this tends not to fit as well for the low $L_{\mathrm{R}}$ end \citep[or the high end in the case of][]{Mazzolari2024}.
While such efforts are informative for linking the X-ray and radio populations, our measurements of the XRLF quantify the space density of X-ray and radio AGN populations over this multi-dimensional parameter space more directly.
We show that with the broad range of radio luminosities possible for a given X-ray luminosity, and vice versa, reconstructing an RLF from an XLF would be a complicated endeavour.

In Figure \ref{RXLF_ratio_Lstar} (left) we show that there is a strong correlation between the fraction of X-ray sources that are radio detected as a function of X-ray luminosity to the total number of sources detected in X-ray (the radio detected fraction) and $L_{\mathrm{X}}/$$L_{\mathrm{*}}$. The characteristic break in the LF (at $L_{\mathrm{*}}$) is due to the combination of the break found in the underlying black hole mass distribution (or galaxy stellar mass function) and the distribution of black hole accretion rates (or Eddington ratios) \citep[e.g.][]{Aird2013, Caplar2015}, which means that sources with Eddington ratio $>1$ are very rare, and this restriction on the highest possible growth rates means that the most luminous ($L>L_{\mathrm{*}}$) and most massive sources are also very rare. 
The higher the mass a black hole has the more likely the source produces a radio jet \citep{Sabater2019, Whittam2022}. High black hole spin has also been suggested as a factor towards the production of a jet \citep[e.g.][]{Blandford1977, Meier2001, Reynolds2021, Whittam2022}, though how this relates with black hole mass is unclear.
At the highest $L_{\mathrm{X}}$ we are seeing the highest (absolute) accretion rates that must be produced by higher mass black holes, thus increased likelihood of a source producing a radio jet. The requirement of a higher mass black hole may explain why in Figure \ref{RXLF_ratio_Lstar} (left) we see the radio AGN fraction of X-ray detected sources reaching 100\% towards high $L_{\mathrm{X}}$. Nevertheless, while radio emission (presumably from a jet) may become a near ubiquitous feature at high $L_{\mathrm{X}}$, it may not necessarily be that luminous, presumably due to differing timescales over which the radiatively efficient accretion (producing the X-ray emission) lasts versus the time needed to produce the extent and thus luminosity of the jet. 

We also present, in Figure \ref{RXLF_ratio_Lstar} (right), a correlation, though weaker, between the fraction of radio sources that are X-ray detected as a function of radio luminosity to the total number of sources detected in radio (the X-ray detected fraction) and $L_{\mathrm{R}}/$$L_{\mathrm{*}}$. At the highest $L_{\mathrm{R}}$ we have the sources that are most likely producing large radio jets, as well as the higher black hole masses as black hole mass has been found to scale with radio luminosity \citep[e.g.][]{Herbert2011}, which links to a higher X-ray detected fraction. Luminous in the radio does not equate to luminous in the X-ray: a possible explanation for this could be that the accretion disk and corona (and therefore the X-ray luminosity) does not persist for the same timescales as a radio jet, and thus may have faded to below the X-ray AGN cut ($L_{\mathrm{X}}>10^{42}$).
Such fading of the coronal emission could also explain the weaker correlation we see for the X-ray detected fraction of radio AGN with increasing $L_\mathrm{R}$ (in comparison to the radio detected fraction of X-ray AGN with increasing $L_\mathrm{X}$).

Our approach benefits from a direct measurement of the space densities of AGN over the space of both $L_{\mathrm{R}}$ and $L_{\mathrm{X}}$. We do not attempt to fit a parametric relation in this work, but if a suitable parameterisation could be found then the space described by the XRLF could be extended to include radio sources that are \emph{not} X-ray detected, and vice versa, thus providing a complete picture of the AGN population and quantifying the links (and lack thereof) between the X-ray and radio luminosities. Furthermore, the same approach could be extended further to incorporate additional wavebands to expand our quantification of the AGN population. The different timescales on which X-ray and radio emission operate make it a challenge to find direct correlations given the difference between observations of emission that can vary over hours/days (X-ray) versus emission that can persist (at the lowest frequencies) up to 10$^8$ years (radio). Adding an additional observation at a wavelength that lies between these timescales could help bridge the gap.

\section{Summary and Conclusions}\label{sec:conc}
We used an X-ray and radio detected sample (radio: \tnum{18,880}; X-ray: \tnum{7491}; X-ray+radio: \tnum{1538}) with optical counterparts and $z>0$ observed in the COSMOS and Bo\"otes fields at 0.5--7~keV in the X-ray and 3~GHz and 144~MHz in the radio, respectively, to investigate the X-ray and radio correlation and measure the multi-dimensional X-ray radio luminosity function across redshift. In this work:
\begin{itemize}
    \item We find a lack of a direct, underlying correlation between the X-ray and radio luminosities of AGN at any redshift. Our results show a broad range of X-ray luminosity for any given radio luminosity (and vice versa) and relations from prior literature appear to be following the sensitivity limits of the surveys.
    \item We modify the \citet{PandC} method for calculating a luminosity function to allow for the space density of sources to be calculated across two different luminosity wave bands at the same time. This method allows us to provide an estimator of the LF that depends on \textit{both} X-ray and radio luminosities and account for the sensitivity across both bands. 
    \item We calculate the space density across luminosity of the overall X-ray and radio samples and find that they agree with parametric models based on prior measurements. However, towards higher redshifts we find that our calculated space densities of sources are under-predicted by the models. 
    \item We calculate the RLF, XLF and XRLF of the X-ray+radio detected sample at different redshifts. We find that for a given X-ray luminosity there is a broad range of radio luminosities, and vice versa. This finding implies that there is no direct correlation between the X-ray and radio luminosities of AGN at any redshift.
    \item We limit the sample to an AGN-dominated sample by using luminosity cuts in X-ray and radio luminosity space. From this we investigated how the fraction of radio-selected AGN within the X-ray sample varies with increasing $L_{\mathrm{X}}$/$L_{\mathrm{*}}$ and how the fraction of X-ray--selected AGN within the radio sample varies with $L_{\mathrm{R}}$/$L_{\mathrm{*}}$.
    We found that, at $L_{\mathrm{X}}>$$L_{\mathrm{*}}$, there is a strong dependence of the radio-selected fraction of X-ray AGN with increasing $L_{\mathrm{X}}$/$L_{\mathrm{*}}$  regardless of redshift. We also see a similar, though weaker, correlation between the X-ray--selected fraction and $L_{\mathrm{R}}$/$L_{\mathrm{*}}$ at $L_{\mathrm{R}}>$$L_{\mathrm{*}}$. This implies that at the brightest X-ray and radio luminosities we are more likely to identify an AGN in both wavebands, however, it will not necessarily be bright in both bands.
\end{itemize}

This work could be extended to include other fields such as ELAIS-S1 \citep{Franzen2015, Ni2021}, eFEDS \citep{Igo2024,Salvato2022} and XMM-LSS \citep{Heywood2020, Chen2018} that benefit from the combination of deep X-ray and radio data. This would increase the number counts for sources across the entire redshift range, improve constraints on the XRLF, and  
allow for the further investigation into the increased fraction of radio-detected X-ray sources towards higher X-ray luminosities (and vice versa). Furthermore, other dimensions could be added to the multi-dimensional luminosity function. For example, the addition of the UV which would trace the accretion disk directly, though this would limit the sample to the unobscured sources where the UV emission can be detected. Lastly, new/ongoing wide-area X-ray and radio surveys \citep[e.g. eROSITA, LOFAR, ASKAP][]{Predehl2021,vanHaarlem2013, Johnston2008} with suitable support from follow-up spectroscopic programmes, as well as new photometric imaging campaigns (such as the Vera Rubin Observatory's optical Legacy Survey of Space and Time) to obtain spectroscopic or photometric redshifts would hugely increase sample sizes. This would increase sample sizes in particular for higher luminosities where our measurements indicate an important connection between X-ray and radio emission may become apparent.


\section*{Acknowledgements}
We thank the anonymous referee for their feedback, which helped improve the paper. C.M.P., J.A. and C.L.B.H acknowledge funding from a UKRI Future Leaders Fellowship (grant codes MR/T020989/1 and MR/Y019539/1). 
For the purpose of open access, the author has applied a Creative Commons Attribution (CC BY) licence to any Author Accepted Manuscript version arising from this submission.

This research has made use the publicly available programming language P\textsc{ython}, including N\textsc{um}P\textsc{y} \citep{Numpy}, S\textsc{ci}P\textsc{y} \citep{Scipy}, M\textsc{atplotlib} \citep{Matplotlib}, A\textsc{stropy} \citep{2013Astropy} and the T\textsc{opcat} analysis program \citep{Taylor2011}.

\section*{Data Availability}

A table of measurements of the XRLF (see Section \ref{XRLFmeas}) for each redshift, along with uncertainties, will be made available alongside the published paper. A Jupyter notebook file will also be made available that will plot the XRLF using the tables provided.



\bibliographystyle{mnras}
\bibliography{ref} 




\appendix

\section{3D X-ray--radio luminosity functions}\label{sec:XRLF}
\begin{figure*}
\centering
    \begin{tabular}{c}
	\includegraphics[width=0.73\textwidth, trim=0.5mm 1mm 0mm 1.5mm, clip]{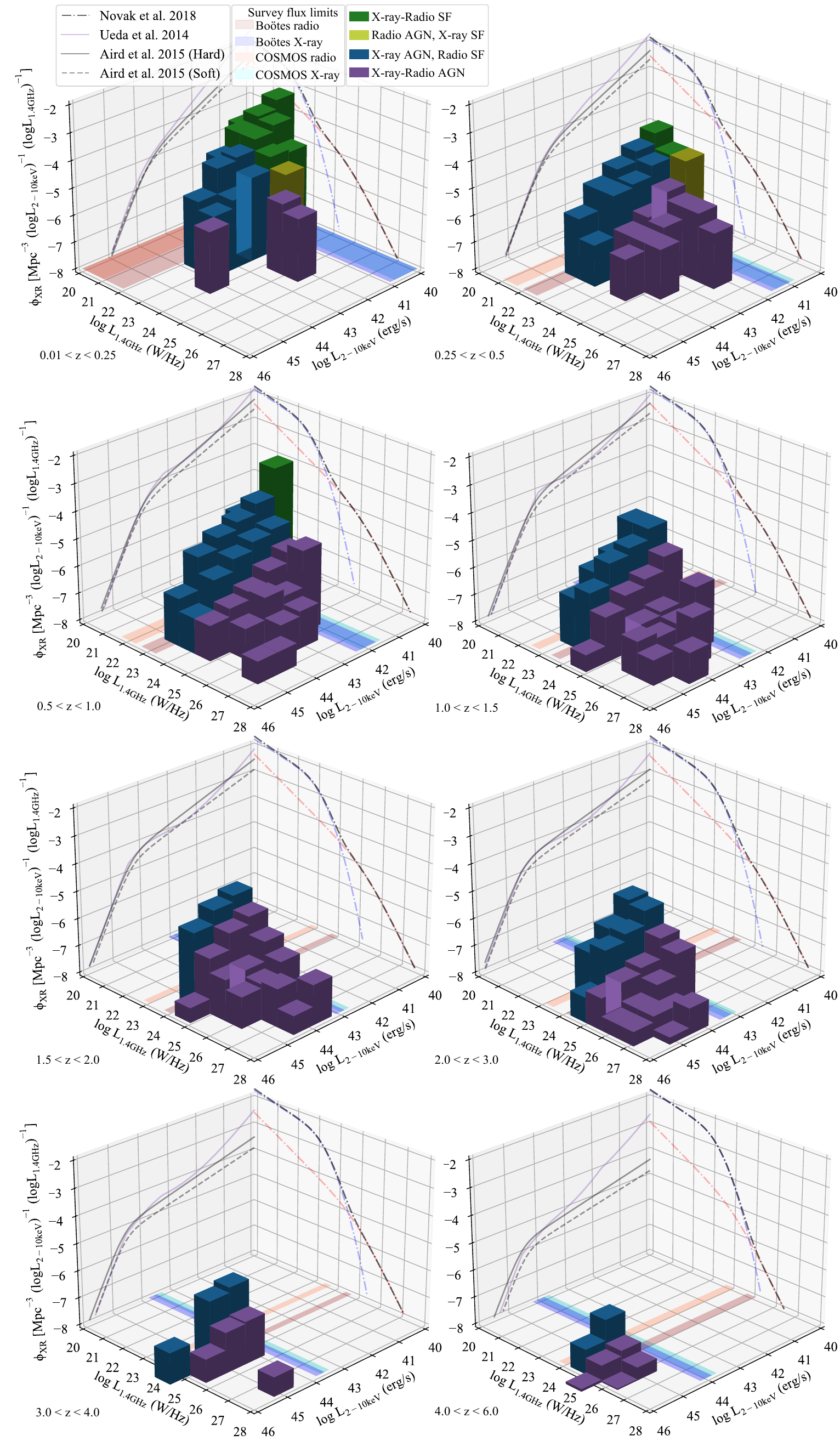}
	\end{tabular}
    \caption{The XRLF, where the density of sources has been calculated per X-ray and radio luminosity bin. The different coloured bins represent the areas of the LF where either SF or AGN are expected to dominate in the radio and/or X-ray bands (see legend for details). The lines on the $\phi_{\mathrm{XR}}$ vs $\log L_{\rm 2 \text{--} 10~keV}$ axis are the XLF models from \citet{Ueda2014, Aird2015} and the lines on the $\phi_{\mathrm{XR}}$ vs $\log L_{\rm 1.4~GHz}$ axis is the RLF model from \citet{Novak2018}, separated into AGN and star-forming sources. The shaded regions on the $\log L_{\rm 1.4~GHz}$ vs $\log L_{\rm 2 \text{--} 10~keV}$ axis represent the flux limits of the X-ray (blue) and radio (red) surveys converted to luminosity space from the beginning to the end of the redshift bins.}
    \label{3DXRLFv}    
\end{figure*}

In this section we present the 3D representations of the XRLF for each redshift range, in Figure \ref{3DXRLFv}. For more information on how these were calculated see Section \ref{sec:MeasLF}. We highlight the sources that most likely have X-ray or radio emission due to star-formation in blue (radio), yellow (X-ray) and green (X-ray and radio). Star-formation due to radio emission can be seen throughout the different redshift ranges, but star-formation due to X-ray emission is not seen at higher redshifts ($z>1$). This is due to the detection limits of the X-ray surveys becoming $L_{\rm 2-10~keV}>10^{42}$~erg/s (the limit used to separate SF from AGN in the X-ray) at higher redshifts.

\section{Probability distribution of R$_{\rm X}$($L_{\mathrm{X}}$, $z$) = log($L_{\rm R}/L_{\rm X}$)}\label{Lafrancasec}

\begin{figure*}
\centering
    \begin{tabular}{c}
	\includegraphics[width=0.9\textwidth, trim=0.5mm 1mm 1mm 0mm, clip]{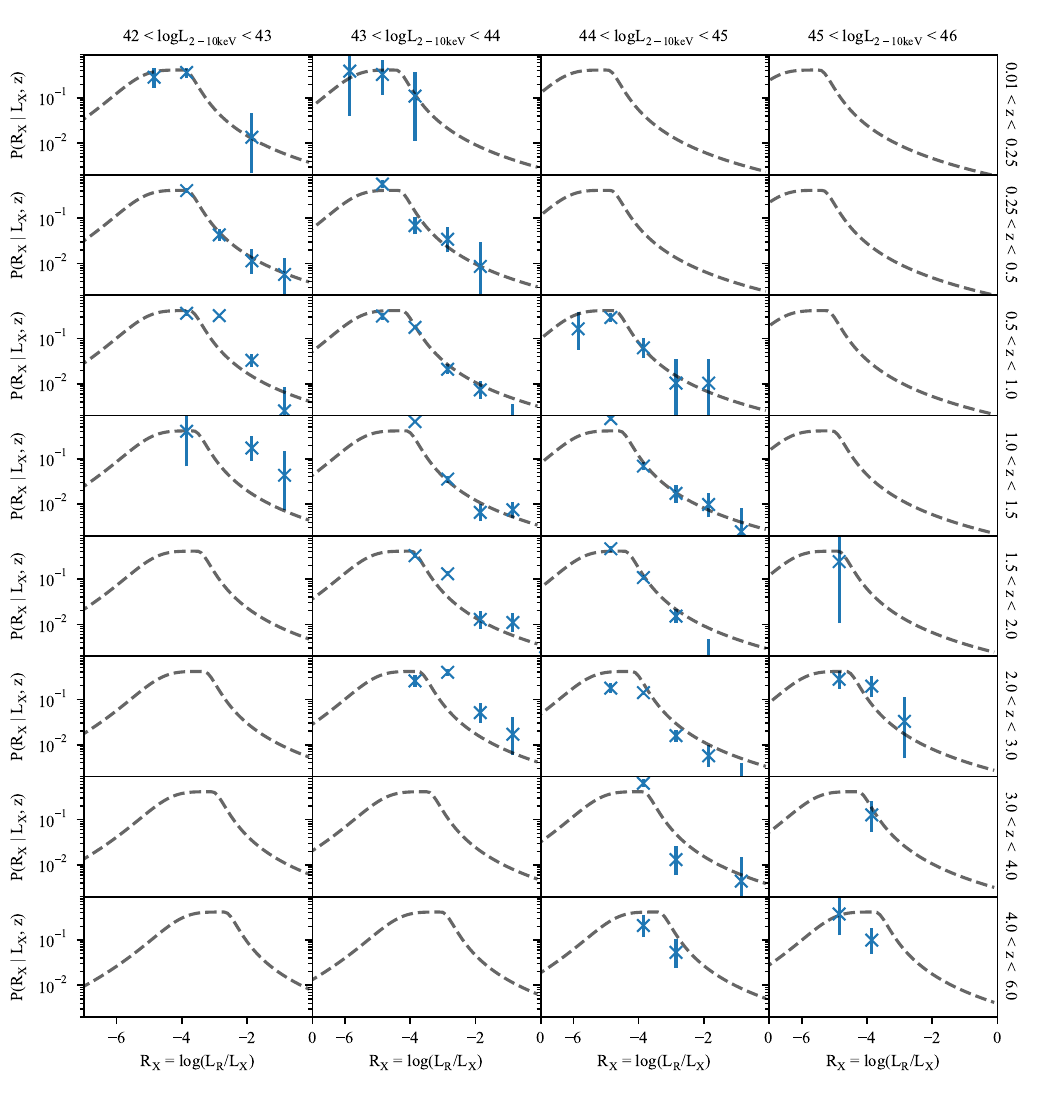}
	\end{tabular}
    \caption{The probability distribution of R$_{\rm X}$ = log(L$_{\rm R}$/L$_{\rm X}$), given L$_{\rm X}$ and $z$. The plot is separated into L$_{\rm X}$ and $z$ bins across the horizontal and vertical directions, respectively. The data from this work are shown as blue crosses and the model from \citet{LaFranca2010} is shown as a dashed grey line. This shows that the data in this work is mostly consistent with the \citet{LaFranca2010} model.}
    \label{Lafrancaplot}    
\end{figure*}

In this section, in Figure \ref{Lafrancaplot}, we present the probability distribution of R$_{\rm X}$($L_{\mathrm{X}}$, $z$) = log($L_{\rm R}/L_{\rm X}$), as discussed in Section \ref{sec:discLF} and first produced in \citet{LaFranca2010}.

\bsp	
\label{lastpage}
\end{document}